\RequirePackage{fix-cm}
\documentclass[a4paper,twocolumn,nopdfoutputerror]{quantumarticle}
\pdfoutput=1
\usepackage[utf8]{inputenc}
\usepackage[english]{babel}
\usepackage[T1]{fontenc}
\usepackage[activate={true,nocompatibility},final,kerning=true,spacing=true,stretch=35,shrink=35]{microtype}
\usepackage{amsmath,amssymb,amsfonts,bm}
\usepackage{graphicx}
\usepackage[numbers,sort&compress]{natbib}
\usepackage{hyperref}
\hypersetup{
    colorlinks=true,
    linkcolor=blue,
    citecolor=blue,
    urlcolor=blue
}
\usepackage{xcolor}
\DeclareRobustCommand{\rev}[1]{#1}
\DeclareRobustCommand{\revb}[1]{#1}
\DeclareRobustCommand{\revg}[1]{#1}

\makeatletter
\let\orig@startsection\@startsection
\def\@startsection#1#2#3#4#5#6{%
  \orig@startsection{#1}{#2}{#3}{#4}{#5}{#6\raggedright}}
\makeatother

\newcommand{\hyph}{-\hspace{0pt}}

\begin{document}

\title{Hardware-Efficient Continuous-Variable Quantum Computation: Finite-$N$ Non-Gaussian Gates in a Floquet-Engineered Cavity-BEC}

\author{Soi-Chan Lei}
\email{d947310@oz.nthu.edu.tw}
\affiliation{Independent Researcher, Macau 999078, China}

\makeatletter
\let\orig@title\@title
\def\@title{\texorpdfstring{\parbox[t]{\linewidth}{\raggedright\orig@title}}{\orig@title}}
\makeatother

\maketitle

\begin{abstract}
We develop a hardware\hyph efficient framework for deterministic continuous\hyph variable (CV) non\hyph Gaussian state and gate synthesis using a single\hyph cavity $^{87}\mathrm{Rb}$ Bose\hyph Einstein condensate (BEC) as a macroscopic bosonic mode, establishing two physical mechanisms and two tensor\hyph network methods. Physically, building on an orthogonal optical dual\hyph driving architecture, we propose unitary Floquet\hyph engineered Two\hyph Axis Twisting (TAT) non\hyph Gaussian dynamics without post\hyph selection, bridging quantum metrology and CV quantum information processing. Through a normal\hyph ordered Holstein\hyph Primakoff expansion and non\hyph perturbative spin simulations, we show that the over\hyph squeezed regime harnesses finite\hyph size Bloch\hyph sphere curvature---manifesting as intrinsic quartic two\hyph photon shearing and Kerr nonlinearities---to deterministically generate Wigner negativity. Methodologically, we formulate an exact automaton Matrix Product Operator in the compact spin basis with an $N$\hyph independent Liouvillian bond dimension ($D_{\mathcal{L}}=6$). This bypasses Fock\hyph space truncation while operating in the full $4^N$ Liouville space without permutation symmetry, accommodating spatially inhomogeneous couplings and site\hyph dependent dissipation where permutational\hyph invariant solvers become inapplicable. Finally, we decouple full\hyph space Lindblad evolution from collective observation via an exact bilateral dynamic programming Dicke projector with step\hyph wise log\hyph scale stabilization, extracting the symmetric\hyph subspace state without moment truncation while tracking symmetry retention. Our $N=50$ open\hyph system benchmarks achieve full\hyph plane Wigner negativity volumes of $1.51$ ($98.8\%$ unconditional state fidelity under uniform relaxation) and $0.76$ (permutation\hyph breaking standing\hyph wave couplings with disorder and dephasing, where PIQS is inapplicable), establishing a pathway toward deterministic non\hyph Gaussian resources in macroscopic BEC bosonic architectures.
\end{abstract}

\section{Introduction}

Synthetic quantum matter provides an exceptionally controllable platform for advancing quantum metrology, nonequilibrium many\hyph body physics, and quantum information processing~\cite{Lei2026,Georgescu2014,Pezze2018,Ritsch2013,Mivehvar2021}. While conventional quantum computing architectures encode logical information in discrete two\hyph level systems (qubits), achieving fault\hyph tolerant operation in discrete\hyph variable architectures demands substantial physical qubit overhead for syndrome extraction. In contrast, continuous\hyph variable (CV) quantum computation utilizing harmonic oscillator modes offers a hardware\hyph efficient paradigm~\cite{Lloyd1999,Joshi2021}. By encoding logical information into the infinite\hyph dimensional Hilbert space of a bosonic mode---such as Gottesman\hyph Kitaev\hyph Preskill (GKP) grid states~\revb{\cite{Gottesman2001,Grimsmo2021}} or Schr\"odinger cat states~\cite{Mirrahimi2014,Ofek2016,Puri2017}---autonomous or hardware\hyph efficient error correction against dominant photon\hyph loss and dephasing channels can be achieved.

Despite its theoretical elegance, universal CV quantum computation is fundamentally bottlenecked by the physical difficulty of synthesizing deterministic, macroscopic non\hyph Gaussian operations. According to the Lloyd\hyph Braunstein criterion~\cite{Lloyd1999} and the Hudson theorem~\cite{Hudson1974}, circuits restricted to Gaussian states, quadratic Hamiltonians, and homodyne measurements maintain non\hyph negative Wigner distributions and can be efficiently simulated on classical computers~\cite{Mari2012,Veitch2012}. Standard optical and microwave resonators readily execute high\hyph fidelity Gaussian gates, but engineering strong higher\hyph order nonlinearities sufficient to deform coherent wavepackets into Wigner\hyph negative states within coherence times remains challenging~\revb{\cite{Grimsmo2021,Puri2017}}. In neutral\hyph atom cavity QED, non\hyph Gaussian entangled states with negative Wigner functions have previously been demonstrated via heralded single\hyph photon detection~\cite{McConnell2015}; however, such measurement\hyph induced protocols are intrinsically probabilistic and rely on post\hyph selection, posing a severe scalability barrier for deterministic quantum circuit execution.

In this work, we overcome these limitations by utilizing a macroscopic single\hyph cavity neutral\hyph atom $^{87}\mathrm{Rb}$ Bose\hyph Einstein condensate (BEC)~\cite{Ritsch2013,Mivehvar2021,Brennecke2007} not merely as an atomic ensemble for metrological spin squeezing~\cite{Kitagawa1993,Leroux2010,Davis2016}, but explicitly as a continuous\hyph variable bosonic \revb{mode for deterministic non\hyph Gaussian resource synthesis}, presenting four main theoretical and computational contributions:

\begin{enumerate}
    \item \textbf{Deterministic Floquet CV Non\hyph Gaussian \revb{Resource Synthesis}:} Building directly upon our microsecond\hyph programmable orthogonal optical dual\hyph driving XXZ architecture~\cite{Lei2026}, rather than relying on low\hyph efficiency probabilistic single\hyph photon heralding and post\hyph selection~\cite{McConnell2015}, we deploy a periodic Floquet sequence of rapid collective $\pi/2$ rotations and \revb{interaction sign inversions}~\cite{Lei2026,Liu2011,Eckardt2017} to dynamically transform native cavity\hyph mediated interactions into a deterministic Two\hyph Axis Twisting (TAT) non\hyph Gaussian generator, steering the macroscopic atomic ensemble into Wigner\hyph negative non\hyph Gaussian states with unit preparation probability.

    \item \textbf{Non\hyph Gaussian \revb{Nonlinearity} from Over\hyph Squeezing and Finite-$N$ Curvature:} Across the existing cavity\hyph BEC and spin\hyph squeezing literature~\cite{Lei2026,Pezze2018,Kitagawa1993,Liu2011}, evolution beyond the optimal Gaussian squeezing time ($t > t_{\mathrm{opt}}$, known as the over\hyph squeezed regime) is universally regarded as a detrimental metrological ceiling because Bloch\hyph sphere wrapping degrades the Wineland spin\hyph squeezing parameter $\xi_R^2$. Through a normal\hyph ordered asymptotic Holstein\hyph Primakoff (H\hyph P) expansion~\cite{Holstein1940,Arecchi1972} complemented by non\hyph perturbative tensor network evolution, we demonstrate that this over\hyph squeezed regime acts as a deterministic non\hyph Gaussian resource: while the $S = N/2 \to \infty$ thermodynamic limit of pure TAT yields a Gaussian two\hyph photon squeezer, the leading\hyph order $\mathcal{O}(1/S)$ geometric curvature of the $\mathrm{SU}(2)$ Bloch sphere reveals the microscopic mechanism driving \rev{quartic two\hyph photon shearing} $a^\dagger a^3 + (a^\dagger)^3 a$ under pure TAT and exact quartic Kerr self\hyph modulation $(a^\dagger a)^2$ under asymmetric Floquet duty cycles~\revb{\cite{Puri2017,Agarwal1997}}, with all higher\hyph order $\mathcal{O}(S^{-2})$ contributions in the deep over\hyph squeezed regime ($\langle\hat{n}\rangle \sim S$) captured non\hyph perturbatively by our microscopic spin simulation \revb{(benchmarked below in the pure TAT limit $\chi_{\mathrm{Kerr}} = 0$)}.

    \item \textbf{Truncation\hyph Free Compact $\mathrm{SU}(2)$ $\mathcal{O}(1)$ Liouvillian MPO Formulation:} Building upon foundational finite\hyph state automaton (FSA) MPO constructions~\cite{Crosswhite2008,McCulloch2007,Hubig2017} and Liouville\hyph space density\hyph matrix vectorization~\cite{Zwolak2004,Verstraete2004,Cui2015,Weimer2021}, we formulate our exact $\mathcal{O}(1)$ Liouvillian superoperator MPO ($D_{\mathcal{L}} = 6$) strictly within the finite collective $\mathrm{SU}(2)$ spin basis. This architecture mathematically eliminates the artificial Fock\hyph space photon\hyph number cutoff $N_{\max}$ that plagues superconducting circuit\hyph QED simulations~\revb{\cite{Joshi2021}}, avoids the phase\hyph coherence breakdown of semiclassical Discrete Truncated Wigner Approximations (DTWA)~\cite{Schachenmayer2015}, \revb{and---unlike permutational\hyph invariant Dicke solvers (PIQS)~\cite{Shammah2018}, which are exact for uniform all\hyph to\hyph all models at $\mathcal{O}(N^3)$ cost---operates directly in the full $4^N$ Liouville space so that spatially inhomogeneous couplings or site\hyph dependent dissipation can be incorporated without symmetry restrictions}.

    \item \textbf{Decoupled Evolution and Observation via an Exact Bilateral Dynamic Programming (DP) Dicke Projector:} While full $4^N$\hyph dimensional Liouville\hyph space \rev{MPO\hyph based exponential time evolution ($W_{\mathrm{II}}$)~\cite{Zaletel2015,Hauschild2018}} is required to capture \revb{general} local single\hyph atom dissipation, reconstructing the Wigner function~\cite{Wigner1932,Dowling1994,Schleich2001} via high\hyph order spin moments $\langle J_x^m J_y^n J_z^k \rangle$ inflates the measurement MPO bond dimension cubically as $(m+1)\allowbreak(n+1)\allowbreak(k+1)$. We construct an Exact Bilateral DP Dicke Projector with \revb{step\hyph wise rescaled} log\hyph gamma anti\hyph overflow normalization that sweeps once across the evolved Liouvillian MPS \rev{$|\rho(t)\rangle\rangle$} to simultaneously track left (ket, $n_L$) and right (bra, $n_R$) excitation counters, extracting the $(N+1) \times (N+1)$ symmetric collective subspace projection $\tilde{\rho}_{n_L,n_R}$ without polynomial moment\hyph truncation loss while rigorously quantifying the symmetric\hyph subspace retention probability $p_{\mathrm{sym}}(t) = \mathrm{Tr}(\tilde{\rho})$.
\end{enumerate}

\section{The Macroscopic Bosonic Mode and Floquet Non-Gaussian Gates}

To establish the neutral\hyph atom Cavity\hyph BEC system as a \revb{deterministic continuous\hyph variable non\hyph Gaussian resource}, we first derive the microscopic Floquet\hyph engineered spin Hamiltonian from our orthogonal optical dual\hyph driving cavity\hyph QED architecture~\cite{Lei2026} and rigorously perform its normal\hyph ordered Holstein\hyph Primakoff mapping onto the bosonic phase plane.

\subsection{Floquet-Engineered Collective Spin Interactions in a Neutral-Atom $^{87}\mathrm{Rb}$ Cavity-BEC}

We build directly upon the microsecond\hyph programmable orthogonal optical dual\hyph driving architecture established in our companion work~\cite{Lei2026}. Concretely, we consider a neutral\hyph atom $^{87}\mathrm{Rb}$ Bose\hyph Einstein condensate~\cite{Ritsch2013,Mivehvar2021,Brennecke2007} of $N$ atoms trapped inside a single high\hyph finesse optical resonator, where the effective pseudospin\hyph $1/2$ states $|\downarrow\rangle, |\uparrow\rangle$ are encoded in two hyperfine ground\hyph state sublevels (such as $|5^2S_{1/2},\allowbreak F=1,\allowbreak m_F=-1\rangle$ and $|F=2,\allowbreak m_F=-2\rangle$, or Raman\hyph coupled Zeeman levels near the $780\,\mathrm{nm}$ $\mathrm{D}_2$ transition~\cite{Brennecke2007,Leroux2010}) with local spin operators $S_k^\alpha = \frac{1}{2}\sigma_k^\alpha$ and collective spin $J_\alpha = \sum_{k=1}^N S_k^\alpha$ ($\alpha \in \{x, y, z\}$). In the far\hyph detuned dispersive cavity\hyph QED regime ($|\Delta_c| \gg \kappa, g\sqrt{N}$), combining a cavity\hyph mediated AC Stark shift (longitudinal density\hyph density coupling $\chi$) with a polarization\hyph orthogonal dual\hyph color Raman driving scheme (transverse flip\hyph flop exchange $\lambda$) synthesizes the exact microscopic lattice Hamiltonian~\cite{Lei2026}:
\begin{equation}
H = \frac{\omega_0}{2}\sum_{k=1}^N \sigma_k^z + \frac{\lambda}{2N}\sum_{k \neq l}^N (\sigma_k^x \sigma_l^x + \sigma_k^y \sigma_l^y) + \frac{\chi}{N}\sum_{k \neq l}^N \sigma_k^z \sigma_l^z.
\label{eq:H_micro}
\end{equation}

Entering the frame rotating at the bare atomic transition frequency $\omega_0$ eliminates the linear Zeeman term $\frac{\omega_0}{2}\sum_{k=1}^N \sigma_k^z$. Projecting Eq.~\eqref{eq:H_micro} onto the collective spin operators via $\sum_{k \neq l} \sigma_k^\alpha \sigma_l^\alpha = 4J_\alpha^2 - N\mathbb{I}$ and invoking the Casimir invariant $J^2 = J_x^2 + J_y^2 + J_z^2$ (which equals $S(S+1)\mathbb{I}$ with $S = N/2$ within the symmetric Dicke manifold) reduces the unmodulated system to the static One\hyph Axis Twisting (OAT) generator $H_{\mathrm{static}} = \frac{4\chi - 2\lambda}{N}J_z^2 + \mathrm{const.}$~\cite{Lei2026}. \revb{Note that although local single\hyph atom dissipation in Sec.~\ref{sec:Liouvillian_MPO} populates lower total\hyph spin sectors ($J < N/2$) where $J^2$ takes distinct eigenvalues $J(J+1)$, for a permutation\hyph invariant initial state and uniform Lindbladian the density matrix $\hat{\rho}(t)$ remains block\hyph diagonal in $J$, so $[J^2, \hat{\rho}(t)] = 0$ holds identically in the full $4^N$ Liouville space and dropping $J^2$ introduces no dynamical error.}

To break the $\mathrm{U}(1)$ rotational symmetry of $H_{\mathrm{static}}$ and synthesize genuine Two\hyph Axis Twisting (TAT) dynamics, we apply our periodic Floquet pulse sequence consisting of rapid global $\pi/2$ rotations around the $x$\hyph axis ($e^{-i(\pi/2)J_x} J_z e^{i(\pi/2)J_x} = -J_y$, which unitarily maps the longitudinal interaction $J_z^2 \to J_y^2$) interspersed with \revb{an optical\hyph drive sign inversion (implemented in the dual\hyph driving cavity\hyph QED setup by switching the dispersive cavity detuning sign $\Delta_c \to -\Delta_c$ or relative two\hyph tone Raman phase between alternating sub\hyph intervals $(\tau_1, \tau_2)$ within each Floquet cycle $T_F = \tau_1 + \tau_2 \ll |\chi|^{-1}, |\lambda|^{-1}$, since an atomic spin rotation alone leaves $J_y^2$ invariant) to invert the sign of the effective transverse interaction block ($+J_y^2 \to -J_y^2$)~\cite{Lei2026} (distinct from asymmetric\hyph pulse OAT\hyph to\hyph TAT schemes without interaction sign inversion~\cite{Liu2011})}. In the fast\hyph pulsing limit ($\omega_F = 2\pi / T_F \gg |\chi|, |\lambda|$), the leading\hyph order Magnus expansion yields the time\hyph averaged Floquet Hamiltonian $H_{\mathrm{Floquet}} = \frac{2}{N}(w_z J_z^2 + w_y J_y^2) = \allowbreak \frac{2}{N}\big[\frac{w_z - w_y}{2}(J_z^2 - J_y^2) + \allowbreak \frac{w_z + w_y}{2}(J_z^2 + J_y^2)\big]$. Crucially, as established in Ref.~\cite{Lei2026}, any uncancelled symmetric residual $(J_z^2 + J_y^2) = J^2 - J_x^2$ reduces via the Casimir operator directly to a longitudinal $-J_x^2$ interaction up to $J^2$. Defining the effective TAT coupling $\chi_{\mathrm{eff}} \equiv \frac{w_z - w_y}{2}$ and the residual longitudinal detuning $\chi_{\mathrm{Kerr}} \equiv -\frac{w_z + w_y}{2}$ yields the effective Hamiltonian up to a Casimir term:
\begin{equation}
H_{\mathrm{eff}} = \frac{2\chi_{\mathrm{eff}}}{N}(J_z^2 - J_y^2) + \frac{2\chi_{\mathrm{Kerr}}}{N}J_x^2.
\label{eq:Heff}
\end{equation}
\revb{Here we adopt the bosonic\hyph rate normalization $2/N = 1/S$ (corresponding to $\chi_{\mathrm{eff}} = \frac{1}{2}\chi_{\mathrm{eff}}^{\mathrm{[1]}}$ relative to the $1/N$ convention in Eq.~(6) of Ref.~\cite{Lei2026}), so that the leading\hyph order squeezing generator in Eq.~\eqref{eq:HTAT_bosonic} carries unit prefactor $\chi_{\mathrm{eff}}(a^2 + (a^\dagger)^2)$.}
For the symmetric $1/2$\hyph duty\hyph cycle Floquet sequence ($\tau_1 = \tau_2$ with exact sign inversion, such that $w_y = -w_z$ and $\chi_{\mathrm{Kerr}} = 0$)~\cite{Lei2026}, writing $J_\alpha^2 = \frac{N}{4}\mathbb{I} + 2\sum_{i<j} S_i^\alpha S_j^\alpha$ (since $(S_i^\alpha)^2 = \frac{1}{4}\mathbb{I}$) shows that the constant diagonal contributions $\frac{N}{4}\mathbb{I}$ from $J_z^2$ and $J_y^2$ cancel identically. Thus, the exact microscopic two\hyph body spin Hamiltonian implemented in our tensor network framework is:
\begin{align}
H_{\mathrm{TAT}} &= \frac{2\chi_{\mathrm{eff}}}{N}(J_z^2 - J_y^2) \nonumber \\
&= \frac{\chi_{\mathrm{eff}}}{N}\sum_{1 \le i < j \le N} (4S_i^z S_j^z - 4S_i^y S_j^y).
\label{eq:HTAT}
\end{align}
The exact correspondence between this ordered\hyph pair representation and the collective generator---together with its dynamical timescale calibration---is verified in Appendix~\ref{app:B}.

\subsection{Holstein-Primakoff Transformation and Bosonic Mapping}

The macroscopic atomic ensemble resides on a compact $\mathrm{SU}(2)$ Bloch sphere of total spin $S = N/2$, whereas a continuous\hyph variable bosonic mode is defined on the flat phase plane $(X, P)$~\cite{Arecchi1972}. Geometrically, the Holstein\hyph Primakoff (H\hyph P) transformation~\cite{Holstein1940} maps the curved Bloch sphere in the neighborhood of its polarized pole onto the tangent bosonic plane.

Specifically, we initialize the ensemble in a coherent spin state (CSS) $|\mathrm{CSS}_x\rangle = \allowbreak |J=S,\allowbreak m_x=S\rangle = \allowbreak \bigotimes_{k=1}^N |+\rangle_{x,k}$ polarized along the $+x$ axis, where $|\pm\rangle_{x,k} = \allowbreak (|\uparrow\rangle_{z,k} \pm \allowbreak |\downarrow\rangle_{z,k})/\sqrt{2}$, corresponding to the bosonic vacuum $|0\rangle_b$. In this standard Hadamard\hyph rotated $S_k^x$ eigenbasis, $S_k^z|\pm\rangle_{x,k} = \frac{1}{2}|\mp\rangle_{x,k}$ acts as the real spin\hyph flip operator while $-S_k^y|\pm\rangle_{x,k} = \pm\frac{i}{2}|\mp\rangle_{x,k}$ forms the right\hyph handed conjugate pair satisfying $[J_z, -J_y] = iJ_x$. Introducing bosonic annihilation and creation operators $a, a^\dagger$ satisfying $[a, a^\dagger] = 1$ with excitation number $\hat{n} = a^\dagger a \in \{0, 1, \dots, 2S\}$, the exact H\hyph P representation along the $+x$ quantization axis is:
\begin{align}
J_x &= S - a^\dagger a = S - \hat{n}, \label{eq:HP_Jx} \\
J_+^{(x)} &\equiv J_z - i J_y = \sqrt{2S}\sqrt{1 - \frac{\hat{n}}{2S}}\,a, \label{eq:HP_Jplus} \\
J_-^{(x)} &\equiv J_z + i J_y = \sqrt{2S}\,a^\dagger\sqrt{1 - \frac{\hat{n}}{2S}}. \label{eq:HP_Jminus}
\end{align}
From Eqs.~\eqref{eq:HP_Jx}--\eqref{eq:HP_Jminus}, the transverse collective spin components $J_z$ and $J_y$ take the exact Hermitian forms in Eqs.~\eqref{eq:Jz_exact} and \eqref{eq:Jy_exact}:
\begin{align}
J_z &= \sqrt{\frac{S}{2}}\left(\sqrt{1 - \frac{\hat{n}}{2S}}\,a + a^\dagger\sqrt{1 - \frac{\hat{n}}{2S}}\right), \label{eq:Jz_exact} \\
J_y &= i\sqrt{\frac{S}{2}}\left(\sqrt{1 - \frac{\hat{n}}{2S}}\,a - a^\dagger\sqrt{1 - \frac{\hat{n}}{2S}}\right). \label{eq:Jy_exact}
\end{align}
Crucially, because $[a, \hat{n}] = a \neq 0$, the operator square root $\sqrt{1 - \hat{n}/(2S)}$ does not commute with $a$ or $a^\dagger$. \rev{Expanding the square root to $\mathcal{O}(S^{-1})$ via $\sqrt{1 - \frac{\hat{n}}{2S}} = 1 - \frac{\hat{n}}{4S} + \mathcal{O}(S^{-2})$} and maintaining strict normal ordering yields:
\begin{align}
J_z &= \sqrt{\frac{S}{2}}\left[a + a^\dagger - \frac{a^\dagger a^2 + (a^\dagger)^2 a}{4S} + \mathcal{O}(S^{-2})\right], \label{eq:Jz_expanded} \\
J_y &= i\sqrt{\frac{S}{2}}\left[a - a^\dagger - \frac{a^\dagger a^2 - (a^\dagger)^2 a}{4S} + \mathcal{O}(S^{-2})\right]. \label{eq:Jy_expanded}
\end{align}
In the asymptotic limit $S \to \infty$ ($\hat{n}/S \to 0$) of Eqs.~\eqref{eq:Jz_expanded} and \eqref{eq:Jy_expanded}, the transverse spin components map directly onto canonical bosonic quadratures $\hat{X} = (a + a^\dagger)/\sqrt{2} = \allowbreak \lim_{S\to\infty} J_z/\sqrt{S}$ and $\hat{P} = (a - a^\dagger)/(i\sqrt{2}) = \allowbreak \lim_{S\to\infty} (-J_y)/\sqrt{S}$ satisfying $[\hat{X}, \hat{P}] = i$.

\subsection{Breaking the Gaussian Barrier: Normal-Ordered Asymptotic $1/S$ Expansion}

To reveal how finite atom number $N = 2S$ breaks the Gaussian barrier, we substitute Eqs.~\eqref{eq:HP_Jx}--\eqref{eq:HP_Jminus} into the collective Floquet Hamiltonian~\eqref{eq:Heff}. First, we note the exact algebraic identity:
\begin{equation}
J_z^2 - J_y^2 = \frac{1}{2}\left[(J_+^{(x)})^2 + (J_-^{(x)})^2\right].
\label{eq:Jz2_Jy2_identity}
\end{equation}
Using $a f(\hat{n}) = f(\hat{n}+1)a$ in Eq.~\eqref{eq:Jz2_Jy2_identity} and noting that $(1 - \frac{\hat{n}}{2S})(1 - \frac{\hat{n}+1}{2S}) = (1 - \frac{2\hat{n}+1}{4S})^2 - \frac{1}{16S^2}$ cancels quadratic $\mathcal{O}(\hat{n}^2/S^2)$ terms identically upon taking the square root, the square of the \revb{bosonic\hyph lowering (spin\hyph $m_x$\hyph raising) operator $J_+^{(x)}$} expands exactly as:
\begin{align}
(J_+^{(x)})^2 &= 2S\sqrt{1 - \frac{\hat{n}}{2S}}\,a\sqrt{1 - \frac{\hat{n}}{2S}}\,a \nonumber \\
&= 2S\sqrt{\left(1 - \frac{\hat{n}}{2S}\right)\left(1 - \frac{\hat{n}+1}{2S}\right)}\,a^2 \nonumber \\
&= 2S\left[1 - \frac{2\hat{n}+1}{4S} - \frac{1}{32S^2} + \mathcal{O}(S^{-3})\right]a^2 \nonumber \\
&= 2S\,a^2 - a^\dagger a^3 - \frac{1}{2}a^2 - \frac{1}{16S}a^2 + \mathcal{O}(S^{-2}).
\label{eq:Jplus_sq}
\end{align}
Combining Eq.~\eqref{eq:Jplus_sq} with its Hermitian conjugate $(J_-^{(x)})^2$ and substituting into $H_{\mathrm{TAT}} = \frac{\chi_{\mathrm{eff}}}{S}(J_z^2 - J_y^2)$ (since $2/N = 1/S$), we obtain the normal\hyph ordered asymptotic $1/S$ bosonic expansion of the pure TAT Hamiltonian (uniformly convergent in the low\hyph excitation regime $\langle\hat{n}\rangle/(2S) \ll 1$):
\begin{align}
H_{\mathrm{TAT}} &= \underbrace{\chi_{\mathrm{eff}}\big(a^2 + (a^\dagger)^2\big)}_{H_{\mathrm{TAT}}^{(0)} \sim \mathcal{O}(S^0 \chi_{\mathrm{eff}})} \nonumber \\
&\quad \underbrace{- \frac{\chi_{\mathrm{eff}}}{2S}\left[a^\dagger a^3 + (a^\dagger)^3 a + \frac{1}{2}\big(a^2 + (a^\dagger)^2\big)\right]}_{H_{\mathrm{TAT}}^{(1)} \sim \mathcal{O}(S^{-1}\chi_{\mathrm{eff}})} \nonumber \\
&\quad + \mathcal{O}(S^{-2}).
\label{eq:HTAT_bosonic}
\end{align}
Furthermore, when the Floquet pulse intervals are tuned to retain a finite longitudinal detuning $\chi_{\mathrm{Kerr}} \neq 0$ in Eq.~\eqref{eq:Heff}, the term $+\frac{\chi_{\mathrm{Kerr}}}{S}J_x^2 = +\frac{\chi_{\mathrm{Kerr}}}{S}(S - \hat{n})^2$ (using $2/N = 1/S$) contributes an exact photon\hyph number Kerr nonlinearity without higher\hyph order $1/S$ corrections:
\begin{equation}
+\frac{\chi_{\mathrm{Kerr}}}{S}J_x^2 = \chi_{\mathrm{Kerr}}S - 2\chi_{\mathrm{Kerr}}a^\dagger a + \frac{\chi_{\mathrm{Kerr}}}{S}(a^\dagger a)^2.
\label{eq:HKerr_bosonic}
\end{equation}

Equations~\eqref{eq:HTAT_bosonic} and \eqref{eq:HKerr_bosonic} characterize two distinct dynamical regimes of the collective spin\hyph boson mapping:
\begin{enumerate}
    \item \textbf{Thermodynamic Gaussian Limit ($S \to \infty$):} Retaining only the leading order $H_{\mathrm{TAT}}^{(0)} = \allowbreak \chi_{\mathrm{eff}}(a^2 + \allowbreak (a^\dagger)^2) = \allowbreak \chi_{\mathrm{eff}}(\hat{X}^2 - \allowbreak \hat{P}^2)$ yields the purely quadratic two\hyph photon squeezing unitary $\hat{U}_0(t) = \allowbreak \exp(-i H_{\mathrm{TAT}}^{(0)} t) = \allowbreak \exp[-i\chi_{\mathrm{eff}}t(a^2 + \allowbreak (a^\dagger)^2)]$. Because the Heisenberg equations of motion $\frac{d}{dt}(\hat{X} \pm \hat{P}) = \allowbreak i[H_{\mathrm{TAT}}^{(0)}, \hat{X} \pm \hat{P}] = \allowbreak \mp 2\chi_{\mathrm{eff}}(\hat{X} \pm \hat{P})$ decouple along the $\pm 45^\circ$ phase\hyph space diagonals, this generator exponentially compresses the variance along the diagonal quadrature $\hat{X}_{\pi/4} = (\hat{X} + \hat{P})/\sqrt{2}$ as $\langle\Delta\hat{X}_{\pi/4}^2(t)\rangle = \frac{1}{2}e^{-4\chi_{\mathrm{eff}}t}$ while stretching the conjugate anti\hyph squeezed quadrature $\hat{P}_{\pi/4} = (\hat{P} - \hat{X})/\sqrt{2}$ as $\frac{1}{2}e^{+4\chi_{\mathrm{eff}}t}$. However, $H_{\mathrm{TAT}}^{(0)}$ alone preserves Gaussianity and cannot generate Wigner negativity~\revb{\cite{Hudson1974,Mari2012}}.

    \item \textbf{Non\hyph Gaussian \revb{State} Generation in the Over\hyph Squeezed Regime ($\langle\hat{n}\rangle \sim S^{1/2}$ to $S$):} In quantum metrology~\cite{Lei2026,Pezze2018,Kitagawa1993,Liu2011}, evolution past the optimal squeezing time $t_{\mathrm{opt}} \sim \frac{\ln N}{4\chi_{\mathrm{eff}}}$ causes the squeezed ellipse to wrap around the Bloch sphere, degrading the Wineland parameter $\xi_R^2$. Here, we show that this over\hyph squeezed regime provides the higher\hyph order nonlinearity required for CV quantum information processing. Under pure TAT ($\chi_{\mathrm{Kerr}} = 0$), the $\Delta n = \pm 2$ parity conservation of $J_z^2 - J_y^2$ forbids diagonal $(a^\dagger a)^2$ terms; instead, Bloch\hyph sphere curvature manifests at $\mathcal{O}(1/S)$ as the \rev{quartic intensity\hyph dependent two\hyph photon shearing term $-\frac{\chi_{\mathrm{eff}}}{2S}(a^\dagger a^3 + (a^\dagger)^3 a) = -\frac{\chi_{\mathrm{eff}}}{2S}[\hat{n}a^2 + (a^\dagger)^2\hat{n}]$ (generating a cubic\hyph amplitude nonlinear force in the Heisenberg equations of motion)}. As squeezing amplifies $\langle\hat{n}\rangle$, this term slows the expansion at the poles of the Bloch sphere, shearing and folding the elongated wavepacket to create deterministic quantum interference fringes without \revb{the probabilistic post\hyph selection required in heralded protocols~\cite{McConnell2015}}. When combined with $\chi_{\mathrm{Kerr}} \neq 0$ in Eq.~\eqref{eq:HKerr_bosonic}, the quartic Kerr interaction $+\frac{\chi_{\mathrm{Kerr}}}{S}(a^\dagger a)^2$ \revb{can drive two\hyph photon Kerr cat\hyph state bifurcations~\cite{Puri2017,Agarwal1997}, provided that the $\mathcal{O}(S)$\hyph larger linear detuning term $-2\chi_{\mathrm{Kerr}}a^\dagger a$ satisfies the parametric instability threshold $|\chi_{\mathrm{Kerr}}| < |\chi_{\mathrm{eff}}|$ or is compensated by an external longitudinal drive $+2\chi_{\mathrm{Kerr}}J_x$. In our open\hyph system numerical benchmark below, we focus on the pure TAT limit ($\chi_{\mathrm{Kerr}} = 0$); including $\chi_{\mathrm{Kerr}} \neq 0$ simply adds a third $S_i^x S_j^x$ automaton channel (increasing $D_W$ from $4$ to $5$ and $D_{\mathcal{L}}$ from $6$ to $8$).}
\end{enumerate}

Crucially, because the Taylor expansion of $\sqrt{1 - \hat{n}/(2S)}$ in Eq.~\eqref{eq:HTAT_bosonic} is an asymptotic series valid for $\langle\hat{n}\rangle/(2S) \ll 1$, higher\hyph order $\mathcal{O}(S^{-2}, S^{-3})$ terms become comparable to $H_{\mathrm{TAT}}^{(1)}$ deep inside the over\hyph squeezed regime ($\langle\hat{n}\rangle \sim S$). Thus, Eq.~\eqref{eq:HTAT_bosonic} serves strictly to elucidate the leading\hyph order physical origin of curvature\hyph induced non\hyph Gaussianity, whereas our rigorous proof of macroscopic Wigner negativity in the deep over\hyph squeezed regime relies on our tensor network simulations, which directly evolve the exact microscopic spin Hamiltonian Eq.~\eqref{eq:HTAT} and non\hyph perturbatively capture all orders of $1/S$.

\section{Exact Finite-State Automaton MPO in Liouville Space}
\label{sec:Liouvillian_MPO}

Simulating highly nonlinear continuous\hyph variable dynamics in conventional Fock\hyph space frameworks requires truncating the bosonic Hilbert space to $\{|0\rangle, |1\rangle, \dots, |N_{\max}\rangle\}$, which suffers from severe cutoff errors when macroscopic cat\hyph like superpositions populate high photon numbers~\revb{\cite{Joshi2021}}. Furthermore, while collective spin dynamics with purely global jump operators remain confined to the $(N+1)$\hyph dimensional symmetric Dicke manifold, realistic open\hyph system cavities experience local single\hyph atom spontaneous emission and dephasing that break permutation symmetry. Existing permutational\hyph invariant Dicke\hyph space solvers (PIQS) capture \revb{uniform} local noise via Clebsch\hyph Gordan blocks with $\mathcal{O}(N^3)$ Liouvillian matrix dimensions~\cite{Shammah2018} \revb{(and cannot treat spatially inhomogeneous couplings or site\hyph dependent disorder)}, whereas semiclassical Discrete Truncated Wigner Approximations (DTWA)~\cite{Schachenmayer2015} fail to resolve delicate high\hyph frequency negative Wigner fringes at late times.

\subsection{Compact $\mathrm{SU}(2)$ Automaton Encoding Without Fock-Space Truncation}

To represent the all\hyph to\hyph all microscopic Hamiltonian Eq.~\eqref{eq:HTAT} on a one\hyph dimensional $N$\hyph site spin\hyph $1/2$ chain without spatial or Fock\hyph space truncation, we adapt the foundational finite\hyph state automaton (FSA) Matrix Product Operator (MPO) formalism \revb{introduced by McCulloch~\cite{McCulloch2007} and Crosswhite and Bacon~\cite{Crosswhite2008} and generalized by Hubig \textit{et al.}~\cite{Hubig2017}}. Because our FSA MPO operates strictly within the finite collective $\mathrm{SU}(2)$ spin basis rather than an infinite bosonic Fock basis, it completely eliminates artificial Fock\hyph space cutoff errors~\revb{\cite{Joshi2021}}.

Specifically, a single all\hyph to\hyph all two\hyph body interaction $\sum_{i<j} c_\alpha S_i^\alpha S_j^\alpha$ requires a 3\hyph state automaton (State 0: no operator applied; State 1: first operator $S_i^\alpha$ applied; State 2: pair completed with $S_j^\alpha$). Because Eq.~\eqref{eq:HTAT} contains two independent non\hyph commuting interaction channels ($\alpha = z$ and $\alpha = y$) that share the same initial state and final accumulation state, the combined operator $H_{\mathrm{TAT}}$ is represented after merging initial and final accumulation states by an exact $4 \times 4$ block\hyph upper\hyph triangular MPO tensor $W^{[k]}$ of minimal bond dimension $D_W = 4$ at each site $k \in \{1, \dots, N\}$ (whereas a single channel requires a $3 \times 3$ block matrix):
\begin{equation}
W^{[k]} = \begin{pmatrix}
\mathbb{I}_k & S_k^z & S_k^y & 0 \\
0 & \mathbb{I}_k & 0 & c_z S_k^z \\
0 & 0 & \mathbb{I}_k & c_y S_k^y \\
0 & 0 & 0 & \mathbb{I}_k
\end{pmatrix}, \quad D_W = 4 = \mathcal{O}(1),
\label{eq:W_MPO}
\end{equation}
where $c_z = +4\chi_{\mathrm{eff}}/N$ and $c_y = -4\chi_{\mathrm{eff}}/N$. Contracting Eq.~\eqref{eq:W_MPO} as $v_L W^{[1]} \allowbreak W^{[2]} \cdots W^{[N]} v_R$ with left boundary vector $v_L = (1, 0, 0, 0)$ and right boundary vector $v_R = (0, 0, 0, 1)^T$ yields the exact microscopic Hamiltonian $H_{\mathrm{TAT}}$ in Eq.~\eqref{eq:HTAT} with a strict $\mathcal{O}(1)$ virtual bond dimension $D_W = 4$, completely independent of atom number $N$.

\subsection{Liouvillian Superoperator MPO for Open Quantum Systems}

To incorporate realistic single\hyph atom dissipative channels, we build upon the vectorized matrix product density operator (MPDO) framework introduced by Zwolak and Vidal~\cite{Zwolak2004} and Verstraete \textit{et al.}~\cite{Verstraete2004} (see also Refs.~\cite{Cui2015,Weimer2021}), vectorizing the $2^N \times 2^N$ density matrix $\hat{\rho}(t)$ via the Choi\hyph Jamio\l{}kowski isomorphism $|\rho\rangle\rangle = \sum_{i,j} \rho_{i,j}|i\rangle\otimes|j\rangle^*$ into a $4^N$\hyph dimensional Liouville space with local physical dimension $d_{\mathcal{L}} = d^2 = 4$ at each lattice site. The Gorini\hyph Kossakowski\hyph Sudarshan\hyph Lindblad (GKSL) master equation~\cite{Lindblad1976,Gorini1976},
\begin{equation}
\frac{d\hat{\rho}}{dt} = -i[H_{\mathrm{TAT}}, \hat{\rho}] + \sum_{k=1}^N \Gamma_{\mathrm{eff}}\mathcal{D}[S_k^-]\hat{\rho},
\label{eq:GKSL}
\end{equation}
with local spin\hyph relaxation dissipator $\mathcal{D}[S_k^-]\hat{\rho} = S_k^-\hat{\rho}S_k^+ - \frac{1}{2}\{S_k^+ S_k^-, \hat{\rho}\}$ acting in the bare atomic basis ($S_k^\pm = S_k^x \pm i S_k^y$), maps onto the linear superoperator evolution $\frac{d}{dt}|\rho\rangle\rangle = \mathcal{L}|\rho\rangle\rangle$:
\begin{equation}
\mathcal{L} = -i(H_{\mathrm{TAT}}\otimes\mathbb{I} - \mathbb{I}\otimes H_{\mathrm{TAT}}^T) + \sum_{k=1}^N \Gamma_{\mathrm{eff}}\hat{\hat{D}}_k,
\label{eq:Liouvillian}
\end{equation}
where the single\hyph site superoperator dissipator acting on site $k$ is:
\begin{equation}
\hat{\hat{D}}_k = S_k^- \otimes (S_k^-)^* - \frac{1}{2}\big(S_k^+ S_k^- \otimes \mathbb{I}_k + \mathbb{I}_k \otimes (S_k^+ S_k^-)^*\big).
\label{eq:Dk_super}
\end{equation}
Because $\hat{\hat{D}}_k$ in Eqs.~\eqref{eq:GKSL}--\eqref{eq:Dk_super} is purely on\hyph site, it embeds directly into the single\hyph site transition slot ($0 \to \mathrm{final}$) of the automaton MPO. By merging the initial and final virtual states of the left\hyph acting Hamiltonian MPO ($-iH_{\mathrm{TAT}}\otimes\mathbb{I}$) and right\hyph acting Hamiltonian MPO ($+i\mathbb{I}\otimes H_{\mathrm{TAT}}^T$), the complete Liouvillian superoperator $\mathcal{L}$ is represented by an exact $6 \times 6$ block\hyph upper\hyph triangular MPO with virtual bond dimension $D_{\mathcal{L}} = 6$ (comprising one initial state, two ket\hyph interaction channels, two bra\hyph interaction channels, and one final accumulation state), strictly $\mathcal{O}(1)$ with respect to $N$.

\revb{We note two physical modeling points regarding Eq.~\eqref{eq:GKSL}. First, for hyperfine ground states of $^{87}\mathrm{Rb}$, direct radiative decay between sublevels is negligible; actual single\hyph atom decoherence in a dispersive cavity\hyph BEC arises from off\hyph resonant Raman and Rayleigh light scattering (generating local dephasing $\mathcal{D}[S_k^z]$ and bidirectional spin flips $\mathcal{D}[S_k^\pm]$) alongside collective cavity photon loss. Second, under rapid Floquet $\pi/2$ rotations around the $x$\hyph axis, bare jump operators transform in the toggling frame into time\hyph averaged combinations of transverse and longitudinal channels. Here, we adopt uniform $\mathcal{D}[S_k^-]$ as a canonical symmetry\hyph breaking single\hyph atom dissipation benchmark that enables direct quantitative cross\hyph validation against exact PIQS~\cite{Shammah2018}. Crucially, because any local Lindblad channel---including toggling\hyph frame\hyph averaged dephasing $\mathcal{D}[S_k^z]$, bidirectional flips $\mathcal{D}[S_k^\pm]$, and spatially inhomogeneous rates $\Gamma_k$---is purely on\hyph site, it embeds into the exact same $(0 \to \mathrm{final})$ MPO slot without increasing $D_{\mathcal{L}} = 6$.}

\rev{Because the dissipative Liouvillian $\mathcal{L}$ is non\hyph Hermitian, real\hyph time open\hyph system evolution is integrated via the second\hyph order $W_{\mathrm{II}}$ MPO exponential time\hyph evolution algorithm of Zaletel \textit{et al.}~\cite{Zaletel2015} implemented in TeNPy~\cite{Hauschild2018} (\texttt{ExpMPOEvolution} with zip\hyph up compression). Specifically, from the block\hyph upper\hyph triangular Liouvillian MPO $\mathcal{L}$, the $W_{\mathrm{II}}$ construction compiles a compact propagator MPO $W_{\mathrm{II}}(\tau) = e^{\mathcal{L}\tau} + \mathcal{O}(\tau^2)$ of virtual bond dimension $D_{\mathcal{L}} - 1 = 5$; composing two complex sub\hyph steps $\tau_{1,2} = \frac{1 \pm i}{2}\Delta t$ cancels the $\mathcal{O}(\Delta t^2)$ error ($\tau_1 + \tau_2 = \Delta t$, $\tau_1^2 + \tau_2^2 = 0$), yielding the second\hyph order update
\begin{equation}
|\rho(t + \Delta t)\rangle\rangle = W_{\mathrm{II}}(\tau_1)\,W_{\mathrm{II}}(\tau_2)\,|\rho(t)\rangle\rangle + \mathcal{O}(\Delta t^3),
\label{eq:WII_step}
\end{equation}
which natively propagates long\hyph range non\hyph Hermitian generators without requiring Hermiticity. A quantitative comparison against existing numerical paradigms is summarized in Table~\ref{tab:comparison}.}

\begin{table*}[!t]
\centering
\caption[]{Quantitative comparison of numerical simulation paradigms for macroscopic Cavity-BEC continuous-variable and spin-squeezing dynamics under local and collective dissipation.}
\label{tab:comparison}
\footnotesize
\renewcommand{\arraystretch}{1.35}
\begin{tabular}{p{2.6cm} p{2.1cm} p{2.5cm} p{1.6cm} p{6.1cm}}
\hline\hline
\textbf{Simulation Framework} & \textbf{Hilbert / Liouville Dim.} & \textbf{Operator Representation} & \textbf{Local Noise?} & \textbf{Key Capabilities \& Bottlenecks} \tabularnewline
\hline
\raggedright Truncated Fock Space~\revb{\cite{Joshi2021}} & \raggedright $(N_{\max}+1)^2$ & \raggedright Dense / Sparse Matrix & \raggedright Approximate & Severe Fock cutoff errors for high\hyph photon cat states; misses exact finite-$N$ $\mathrm{SU}(2)$ boundary curvature. \tabularnewline
\raggedright Collective Dicke Basis & \raggedright $(N+1)^2$ & \raggedright Symmetric $\mathrm{SU}(2)$ Matrix & \raggedright No (Global only) & Exact for unitary and collective cavity decay; fails when local spontaneous emission breaks permutation symmetry. \tabularnewline
\raggedright Permutational Invariant (PIQS)~\cite{Shammah2018} & \raggedright $\mathcal{O}(N^3)$ Dicke Blocks & \raggedright Block\hyph Diagonal Liouvillian & \raggedright Yes (Uniform only) & Exact for uniform local dissipation via Clebsch\hyph Gordan blocks ($\sim N^3/6$ memory); \revb{inapplicable when spatial inhomogeneity or disorder breaks permutation symmetry}. \tabularnewline
\raggedright Semiclassical DTWA~\cite{Schachenmayer2015} & \raggedright $\mathcal{O}(N)$ Trajectories & \raggedright Monte Carlo Sampling & \raggedright Approximate & Efficient for short\hyph time squeezing ($\xi_R^2$); fails to resolve fragile high\hyph frequency Wigner negativity at late times. \tabularnewline
\raggedright \textbf{FSA Liouvillian MPO + Bilateral DP (This Work)} & \raggedright $4^N$ (Compressed to $\chi_{\mathrm{MPS}}$) & \raggedright Exact $\mathcal{O}(1)$ Automaton ($D_{\mathcal{L}}=6$) & \raggedright \revb{Yes (General)} & Truncation\hyph free MPO construction; decouples full\hyph space \rev{$W_{\mathrm{II}}$ Lindblad} evolution from moment\hyph truncation\hyph free symmetric\hyph subspace projection ($\tilde{\rho}_{n_L,n_R}$) and tracks $p_{\mathrm{sym}}(t)$; \revb{extends natively to inhomogeneous couplings/dissipation (limited by $\chi_{\mathrm{MPS}}$ growth at deep over\hyph squeezing)}. \tabularnewline
\hline\hline
\end{tabular}
\end{table*}

\section{Wigner Negativity Reconstruction and Dynamic Programming Projection}

\subsection{Phase-Space Wigner Function and Negativity Metrics}

To certify that the Floquet\hyph driven Cavity\hyph BEC operates in the non\hyph Gaussian regime required for universal CV quantum computation~\cite{Mari2012,Veitch2012}, we evaluate the continuous\hyph variable Wigner quasiprobability distribution $W(X, P)$~\cite{Wigner1932,Schleich2001}:
\begin{equation}
W(X, P) = \frac{1}{\pi}\int_{-\infty}^\infty e^{2iPY}\langle X - Y|\hat{\rho}_{\mathrm{boson}}(t)|X + Y\rangle\,dY,
\label{eq:Wigner_CV}
\end{equation}
where $\iint_{-\infty}^\infty W(X, P)\,dX\,dP = 1$ and $|W(X, P)| \le \allowbreak 1/\pi \approx \allowbreak 0.3183$ (in units where $\hbar = 1$ and $[\hat{X}, \hat{P}] = i$). Equivalently, on the compact $\mathrm{SU}(2)$ Bloch sphere, the intrinsic spherical Wigner distribution $W_S(\theta, \phi)$ is defined via the Stratonovich\hyph Agarwal\hyph Dowling\hyph Schleich multipole expansion~\cite{Dowling1994}:
\begin{align}
W_S(\theta, \phi) &= \sqrt{\frac{2S+1}{4\pi}}\sum_{K=0}^{2S}\sum_{Q=-K}^K \mathrm{Tr}\big(\hat{\rho}_S(t)\hat{T}_{KQ}^{(S)\dagger}\big) \nonumber \\
&\quad \times Y_{KQ}(\theta, \phi),
\label{eq:Wigner_SU2}
\end{align}
\revb{where $\hat{T}_{KQ}^{(S)}$ denote the orthonormal irreducible spherical tensor (multipole) operators satisfying $\mathrm{Tr}(\hat{T}_{KQ}^{(S)\dagger}\hat{T}_{K'Q'}^{(S)}) = \delta_{KK'}\delta_{QQ'}$ and $Y_{KQ}(\theta, \phi)$ are standard spherical harmonics~\cite{Dowling1994}.} Near the polarized pole ($\langle\hat{n}\rangle \ll 2S$), Eq.~\eqref{eq:Wigner_SU2} converges under tangent\hyph plane (or stereographic) H\hyph P projection around $(\theta_0 = \pi/2, \phi_0 = 0)$ directly to Eq.~\eqref{eq:Wigner_CV} \rev{up to the area\hyph element Jacobian $S^{-1}$ ($S^{-1}W_S(\theta, \phi) \to W(X, P)$ as $dX\,dP = S\,d\Omega$)}~\cite{Dowling1994}. \revb{Deep inside the over\hyph squeezed regime ($\langle\hat{n}\rangle \sim S$), where the state wraps across the Bloch sphere and the local tangent\hyph plane approximation breaks down, Eq.~\eqref{eq:Wigner_CV} evaluated via the exact Fock identification $|J=S, m_x=S-n\rangle \leftrightarrow |n\rangle_b$ represents the Fock\hyph mapped planar bosonic Wigner function, whereas Eq.~\eqref{eq:Wigner_SU2} (likewise directly computable from the extracted $(N+1)\times(N+1)$ matrix $\rho_{n_L,n_R}$) captures the intrinsic compact $\mathrm{SU}(2)$ phase\hyph space geometry.} We quantify non\hyph classical phase\hyph space interference via two rigorous metrics~\cite{Kenfack2004}: (i) the global Wigner minimum $W_{\min} = \min_{X,P} W(X, P) < 0$, and (ii) the integrated Wigner negativity volume $\delta_{\mathcal{W}}(t)$ in Eq.~\eqref{eq:Wigner_volume}:
\begin{equation}
\delta_{\mathcal{W}}(t) = \iint_{-\infty}^\infty \big(|W(X, P, t)| - W(X, P, t)\big)\,dX\,dP \ge 0.
\label{eq:Wigner_volume}
\end{equation}
\revb{Because the integrand $|W| - W = 2\max(0, -W) \ge 0$ is everywhere non\hyph negative, integrating Eq.~\eqref{eq:Wigner_volume} over any finite phase\hyph space window $\Omega$ yields a strict conservative lower bound $\delta_{\mathcal{W}}^{\Omega}(t) \le \delta_{\mathcal{W}}(t)$ on the full\hyph plane negativity volume.}

\subsection{Decoupling Evolution from Observation via Exact Bilateral Dynamic Programming}
\label{sec:DP}

While the $\mathcal{O}(1)$ Liouvillian MPO circumvents Fock\hyph space truncation during real\hyph time evolution, reconstructing $W(X, P)$ or $W_S(\theta, \phi)$ up to multipole rank $K = 2S = N$ directly from macroscopic spin moments $\langle J_x^m J_y^n J_z^k \rangle$ suffers from severe numerical conditioning and a cubic MPO bond\hyph dimension explosion $D_W = (m+1)\allowbreak(n+1)\allowbreak(k+1) = \mathcal{O}(N^3)$, which acts as a low\hyph pass filter that washes out high\hyph frequency Wigner negativity (see Appendix~\ref{app:A} for the exact FSA construction of high\hyph order spin multipoles).

To resolve this observational bottleneck, we formulate an Exact Bilateral Dynamic Programming (DP) Dicke Projection Protocol that algorithmically decouples the $4^N$\hyph dimensional open\hyph system \rev{$W_{\mathrm{II}}$ Lindblad} evolution from the collective phase\hyph space reconstruction. After evolving the vectorized density matrix \rev{$|\rho(t)\rangle\rangle$} in the full $4^N$ Liouville space, we project the state onto the symmetric collective Dicke basis $\{|J=S, m_x=S-n\rangle\}_{n=0}^N$ (labeled directly by the bosonic excitation number $n = S - m_x \in \{0, 1, \dots, N\}$) to extract the unnormalized $(N+1) \times (N+1)$ collective projection matrix in Eq.~\eqref{eq:rho_tilde}:
\begin{align}
\tilde{\rho}_{n_L,n_R}(t) &= \mathrm{Tr}\big(|J, S-n_R\rangle\langle J, S-n_L|\rev{\hat{\rho}(t)}\big) \nonumber \\
&= \langle\langle \Pi_{n_L,n_R}\rev{|\rho(t)\rangle\rangle},
\label{eq:rho_tilde}
\end{align}
where $|J, S-n\rangle = {\binom{N}{n}}^{-1/2} \allowbreak \sum_{\mathrm{perm}, |s|=n} |s_1, \dots, s_N\rangle_x$.

Rather than constructing $(N+1)^2$ separate projector MPOs or performing iterative SVD compressions~\cite{Schollwock2011}, our Bilateral DP Automaton sweeps once across the $N$\hyph site MPS representation of \rev{$|\rho(t)\rangle\rangle$}. \rev{Expressed explicitly in the Hadamard\hyph rotated local $S_k^x$ eigenbasis $\{|s_L, s_R\rangle\rangle_k \equiv |s_L\rangle_{x,k}\otimes|s_R\rangle_{x,k}^*\}$ (with $s_L, s_R \in \{0, 1\}$ indexing the local $|+\rangle_{x,k}, |-\rangle_{x,k}$ states), the evolved state takes the open\hyph boundary MPS form
\begin{align}
|\rho(t)\rangle\rangle &= \sum_{\{s_{L,k}, s_{R,k}\}} \sum_{\{\alpha_k\}} \left(\prod_{k=1}^N M_{\alpha_{k-1},\alpha_k}^{[k](s_{L,k}, s_{R,k})}\right) \nonumber \\
&\quad \times \bigotimes_{k=1}^N |s_{L,k}, s_{R,k}\rangle\rangle_k,
\label{eq:rho_MPS}
\end{align}
where $M_{\alpha_{k-1},\alpha_k}^{[k](s_{L,k}, s_{R,k})}$ is the local MPS tensor of $|\rho(t)\rangle\rangle$ at site $k$ in the $S_k^x$ eigenbasis, and $\alpha_k \in \{1, \dots, \chi_k\}$ are the virtual bond indices with boundary conditions $\alpha_0 = \alpha_N = 1$.} At site $k$, the bilateral counter tracks both the left (ket) excitation count $n_L \in \{0, \dots, k\}$ and the right (bra) excitation count $n_R \in \{0, \dots, k\}$ simultaneously. \revb{In an unrescaled contraction over all permutations of Eq.~\eqref{eq:rho_MPS}, the accumulated amplitude $E_1^{[N]}(n_L, n_R) \equiv \sum_{|s_L|=n_L,\,|s_R|=n_R}\langle\langle s_L, s_R|\rho(t)\rangle\rangle = \sqrt{\binom{N}{n_L}\binom{N}{n_R}}\,\tilde{\rho}_{n_L,n_R}$ grows combinatorially and overflows IEEE\hyph 754 double precision for central excitations at $N > 1029$ (while the standalone normalization factor ${\binom{N}{n}}^{-1}$ underflows to zero). To eliminate this instability, we introduce the step\hyph wise rescaled bilateral tensor $\widetilde{E}_{\alpha_k}^{[k]}(n_L, n_R)$ and cumulative log\hyph scale $\Lambda_k$, updated at every site $k$ via:}
\begin{align}
\revb{U_{\alpha_k}^{[k]}(n_L, n_R)} &= \sum_{\alpha_{k-1}, s_L, s_R} \revb{\widetilde{E}_{\alpha_{k-1}}^{[k-1]}(n_L - s_L, n_R - s_R)} \nonumber \\
&\quad \times M_{\alpha_{k-1},\alpha_k}^{[k](s_L, s_R)}, \label{eq:DP_recursion} \\
\revb{\widetilde{E}_{\alpha_k}^{[k]}(n_L, n_R)} &\revb{{}= \frac{U_{\alpha_k}^{[k]}(n_L, n_R)}{c_k}, \quad \Lambda_k = \Lambda_{k-1} + \ln c_k,} \label{eq:DP_step_rescale}
\end{align}
\rev{with open\hyph boundary initialization \revb{$\widetilde{E}_1^{[0]}(n_L, n_R) = \delta_{n_L,0}\delta_{n_R,0}$}, \revb{$\Lambda_0 = 0$,} and \revb{$\widetilde{E}_{\alpha_k}^{[k]}(n_L, n_R) = 0$} for $n_L < 0$ or $n_R < 0$,} \revb{where $c_k = \max_{\alpha_k, n_L, n_R} |U_{\alpha_k}^{[k]}(n_L, n_R)|$ keeps all intermediate tensors $\mathcal{O}(1)$. At site $k = N$, the accumulated log\hyph scale $\Lambda_N$ restores the exact overall scale factor $e^{\Lambda_N}$ while canceling the combinatorial log\hyph gamma factor directly inside the exponential, achieving an exact, lossless projection without overflow, underflow, or approximation overhead:}
\begin{align}
\tilde{\rho}_{n_L,n_R}(t) &= \revb{\widetilde{E}_1^{[N]}(n_L, n_R)} \nonumber \\
&\quad \times \exp\left\{\revb{\Lambda_N} - \frac{1}{2}\left[\ln\binom{N}{n_L} + \ln\binom{N}{n_R}\right]\right\},
\label{eq:log_gamma}
\end{align}
\revb{with $\ln\binom{N}{n} = \ln\Gamma(N+1) - \ln\Gamma(n+1) - \ln\Gamma(N-n+1)$.}

Physically, while our Bilateral DP algorithm is mathematically lossless within the $J = N/2$ symmetric Dicke subspace (eliminating any polynomial moment\hyph truncation error), local single\hyph atom dissipation $\mathcal{D}[S_k^-]$ inevitably leaks a population fraction $1 - p_{\mathrm{sym}}(t) = 1 - \mathrm{Tr}(\tilde{\rho}) \sim \mathcal{O}(N\Gamma_{\mathrm{eff}}t)$ into non\hyph symmetric subspaces ($J < N/2$) that are projected out by $\Pi_{n_L,n_R}$. We therefore normalize the conditional symmetric\hyph subspace density matrix via $\rho_{n_L,n_R}(t) = \tilde{\rho}_{n_L,n_R}(t) / p_{\mathrm{sym}}(t)$ while explicitly recording $p_{\mathrm{sym}}(t) = \mathrm{Tr}[\tilde{\rho}(t)]$ as a quantitative diagnostic of collective symmetry retention\revb{; this conditional state faithfully represents the unconditional deterministic output whenever $1 - p_{\mathrm{sym}}(t) \ll 1$ (since $1 - p_{\mathrm{sym}}(t)$ directly bounds the weight of the discarded non\hyph symmetric sectors)}. Under the H\hyph P correspondence $|J=S, m_x=S-n\rangle \leftrightarrow |n\rangle_b$, the \revb{Fock\hyph mapped planar} Wigner function Eq.~\eqref{eq:Wigner_CV} is then evaluated analytically in terms of generalized Laguerre polynomials $L_n^{(m-n)}(2R^2)$ ($R^2 = X^2 + P^2$, $\Phi = \mathrm{atan2}(P, X)$)~\cite{Schleich2001} via Eq.~\eqref{eq:Wigner_Laguerre}:
\begin{align}
W(X, P) &= \frac{e^{-R^2}}{\pi}\sum_{n=0}^N (-1)^n \Bigg[\rho_{n,n} L_n^{(0)}(2R^2) \nonumber \\
&\quad + 2\sum_{m=n+1}^N \sqrt{\frac{n!}{m!}}(2R^2)^{\frac{m-n}{2}} L_n^{(m-n)}(2R^2) \nonumber \\
&\quad \times \mathrm{Re}\big(\rho_{m,n} e^{-i(m-n)\Phi}\big)\Bigg].
\label{eq:Wigner_Laguerre}
\end{align}

\begin{figure*}[!t]
\centering
\includegraphics[width=\textwidth]{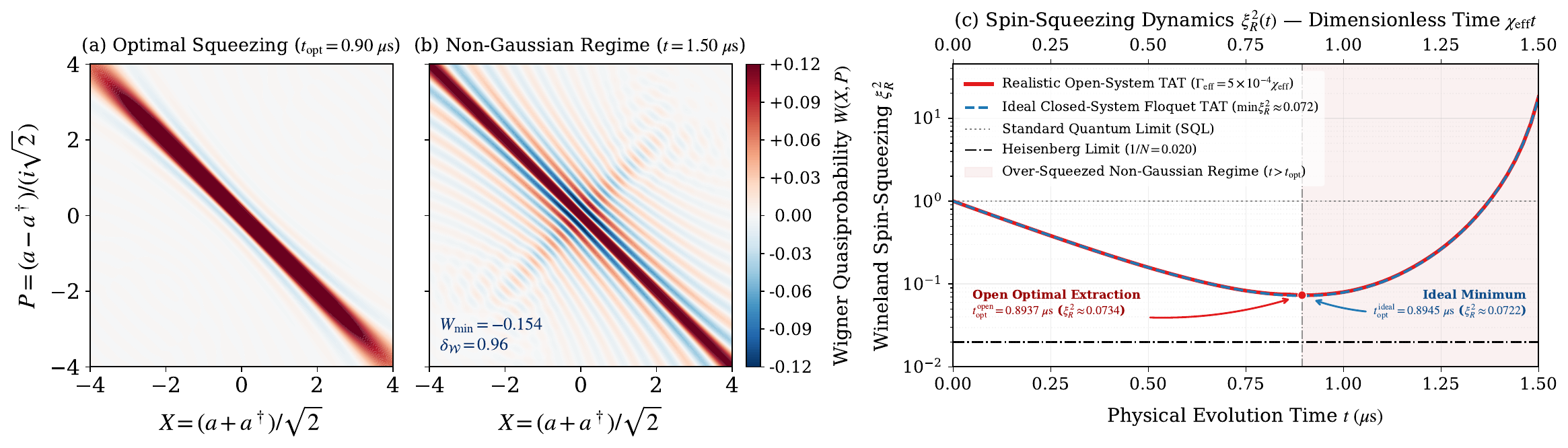}
\caption[]{Phase-space Wigner quasiprobability reconstruction and spin-squeezing dynamics for a proof-of-principle $^{87}\mathrm{Rb}$ ensemble of $N = 50$ atoms ($4^{50} \approx 1.27 \times 10^{30}$ Liouville-space dimensions) under open-system Floquet TAT evolution ($\chi_{\mathrm{eff}} = 1.0\,\mu\mathrm{s}^{-1}$, effective local relaxation rate $\Gamma_{\mathrm{eff}}/\chi_{\mathrm{eff}} = 5 \times 10^{-4}$). (a) Near the exact continuous-time open-system optimal metrological spin-squeezing time \rev{$t_{\mathrm{opt}}^{\mathrm{open}} = 0.8937\,\mu\mathrm{s}$, evaluated at the nearest $\Delta t = 0.02\,\mu\mathrm{s}$ evolution grid point $t = 0.90\,\mu\mathrm{s}$ ($\xi_R^2 \approx 0.0734$, $-11.34\,\mathrm{dB}$, $p_{\mathrm{sym}} \approx 0.9942$)}, the distribution is compressed along the diagonal into a smooth elongated ellipse dominated by the quadratic Gaussian squeezing generator $H_{\mathrm{TAT}}^{(0)}$ [Eq.~\eqref{eq:HTAT_bosonic}]. (b) Inside the over-squeezed regime at \rev{$t = 1.50\,\mu\mathrm{s}$ ($\chi_{\mathrm{eff}}t = 1.50$)}, finite-$N$ Bloch-sphere curvature shears and folds the wavepacket, producing pronounced quantum interference fringes and deep Wigner negativity \rev{($W_{\min} \approx -0.154$, central peak $W_{\max} \approx 0.317$, \revb{windowed} negativity volume $\delta_{\mathcal{W}}^{[-4,4]^2} \approx 0.96$ \revb{over the displayed $[-4,4]^2$ domain, providing a strict lower bound on the converged open-system full-plane volume $\delta_{\mathcal{W}}^{\mathrm{full}} \approx 1.51$ shown in Fig.~\ref{fig:spherical_wigner}(a)}; shared color scale saturated at $\pm 0.12$ to resolve secondary negative interference ripples)}. (c) Dual-axis time evolution of the Wineland spin-squeezing parameter $\xi_R^2(t)$ plotted against physical evolution time $t$ (bottom axis, in $\mu\mathrm{s}$) and dimensionless interaction time $\chi_{\mathrm{eff}}t$ (top axis), comparing the realistic dissipative open-system \rev{$W_{\mathrm{II}}$ trajectory (solid red curve, adaptive $\chi_{\max} = 128 \to 256$, discrete-grid minimum $\xi_R^2 \approx 0.0734$ at the nearest grid point $t = 0.90\,\mu\mathrm{s}$ and SQL crossing at $t \approx 1.372\,\mu\mathrm{s}$)} against the ideal closed-system Floquet TAT benchmark (dashed blue curve, minimum \rev{$\xi_R^2 \approx 0.0722$ at $t_{\mathrm{opt}}^{\mathrm{ideal}} = 0.8945\,\mu\mathrm{s}$}), relative to the Standard Quantum Limit ($\mathrm{SQL} = 1$) and Heisenberg Limit ($1/N = 0.020$). \rev{In continuous time, local relaxation advances the open-system turnaround slightly earlier than the unitary benchmark ($t_{\mathrm{opt}}^{\mathrm{open}} = 0.8937\,\mu\mathrm{s} < t_{\mathrm{opt}}^{\mathrm{ideal}} = 0.8945\,\mu\mathrm{s}$, a $\approx 0.8\,\mathrm{ns}$ shift, both resolving onto $t = 0.90\,\mu\mathrm{s}$ on the $\Delta t = 0.02\,\mu\mathrm{s}$ grid).} The vertical dash-dotted line at $t = t_{\mathrm{opt}}$ demarcates the transition from the Gaussian squeezing regime ($t \le t_{\mathrm{opt}}$) to the over-squeezed non-Gaussian regime ($t > t_{\mathrm{opt}}$, shaded).}
\label{fig:wigner}
\end{figure*}

\subsection{Numerical Benchmark: Proof-of-Principle Demonstration at $N = 50$ and Macroscopic Scaling}

Because our finite\hyph state automaton Liouvillian MPO maintains a strictly $N$\hyph independent virtual bond dimension ($D_{\mathcal{L}} = 6$) and our \revb{step\hyph wise rescaled} log\hyph gamma recursion Eqs.~\eqref{eq:DP_recursion}--\eqref{eq:log_gamma} avoids overflow of $\binom{N}{n}$ for $N > 1029$, the operator and projector representations impose no inherent atom\hyph number ceiling. \revb{In the low\hyph excitation metrological regime ($t \le t_{\mathrm{opt}}$, where $\chi_{\mathrm{MPS}}$ remains modest and the DP counter can be restricted to $n_L, n_R \le n_{\mathrm{cut}} \ll N$), this framework scales directly to macroscopic ensembles (e.g., \revb{as demonstrated for pure\hyph state TDVP at $N = 10^3$ in Ref.~\cite{Lei2026}}), whereas deep inside the over\hyph squeezed regime ($\langle\hat{n}\rangle \sim S = N/2$) both the populated excitation window and the Liouville\hyph space bond dimension $\chi_{\mathrm{MPS}}$ grow polynomially with $N$.} To provide a high\hyph resolution proof\hyph of\hyph principle demonstration of full $4^N$\hyph dimensional open\hyph system evolution deep into the over\hyph squeezed non\hyph Gaussian regime ($t > t_{\mathrm{opt}}$) without visual fringe aliasing, we benchmark an ensemble of $N = 50$ $^{87}\mathrm{Rb}$ atoms ($S = 25$, spanning $4^{50} \approx 1.27 \times 10^{30}$ Liouville\hyph space dimensions, far beyond the $N \sim 20$ limit of exact diagonalization). 
The system is initialized in $|\mathrm{CSS}_x\rangle$ under the microscopic Floquet TAT Hamiltonian Eq.~\eqref{eq:HTAT} with effective coupling rate $\chi_{\mathrm{eff}} = 1.0\,\mu\mathrm{s}^{-1}$ ($\chi_{\mathrm{eff}}/2\pi \approx 159\,\mathrm{kHz}$) and realistic cavity\hyph QED local dissipation rate $\Gamma_{\mathrm{eff}} = 5 \times 10^{-4}\chi_{\mathrm{eff}}$~\cite{Brennecke2007,Leroux2010}, \rev{evolved via second\hyph order $W_{\mathrm{II}}$ MPO propagation with time step $\Delta t = 0.02\,\mu\mathrm{s}$ and an adaptive Liouville\hyph space bond dimension ($\chi_{\max} = 128$ for $t \le 0.86\,\mu\mathrm{s}$ and $\chi_{\max} = 256$ for $0.86\,\mu\mathrm{s} < t \le 1.50\,\mu\mathrm{s}$, maintaining full\hyph space trace conservation $|\mathrm{Tr}\,\hat{\rho} - 1| < 1.3 \times 10^{-3}$, monotonic purity decay to $\mathrm{Tr}\,\hat{\rho}^2 = 0.9771$, and symmetric\hyph subspace retention $p_{\mathrm{sym}} \approx 0.9903$ at $t = 1.50\,\mu\mathrm{s}$)}. To bridge experimental timescales and universal spin\hyph squeezing scaling, Fig.~\ref{fig:wigner}(c) displays both the physical evolution time $t$ (bottom axis) and the dimensionless interaction time $\chi_{\mathrm{eff}}t$ (top axis) for the Wineland spin\hyph squeezing parameter~\revb{\cite{Pezze2018,Wineland1994}},
\begin{equation*}
\xi_R^2(t) \equiv \frac{N \min_{\mathbf{n}_\perp} \langle \Delta J_{\mathbf{n}_\perp}^2(t)\rangle}{|\langle\mathbf{J}(t)\rangle|^2} \rev{\approx \frac{N \langle \Delta (J_z - J_y)^2(t)\rangle}{2\langle J_x(t)\rangle^2}},
\end{equation*}
\rev{where the second relation is exact in the unitary limit (since $\hat{X}_{\pi/4} = (\hat{X}+\hat{P})/\sqrt{2} = \lim_{S\to\infty}(J_z - J_y)/\sqrt{2S}$) and remains accurate to within $10^{-5}$ near $t_{\mathrm{opt}}$ under weak local relaxation.} Specifically, while $t$ denotes the continuous evolution time variable, we define the optimal spin\hyph squeezing extraction time $t_{\mathrm{opt}} \equiv \arg\min_t \xi_R^2(t)$ as the critical turnaround instant at which the Wineland parameter reaches its global minimum before rebounding upward into the over\hyph squeezed regime ($t > t_{\mathrm{opt}}$).

\begin{figure*}[!t]
\centering
\includegraphics[width=\textwidth]{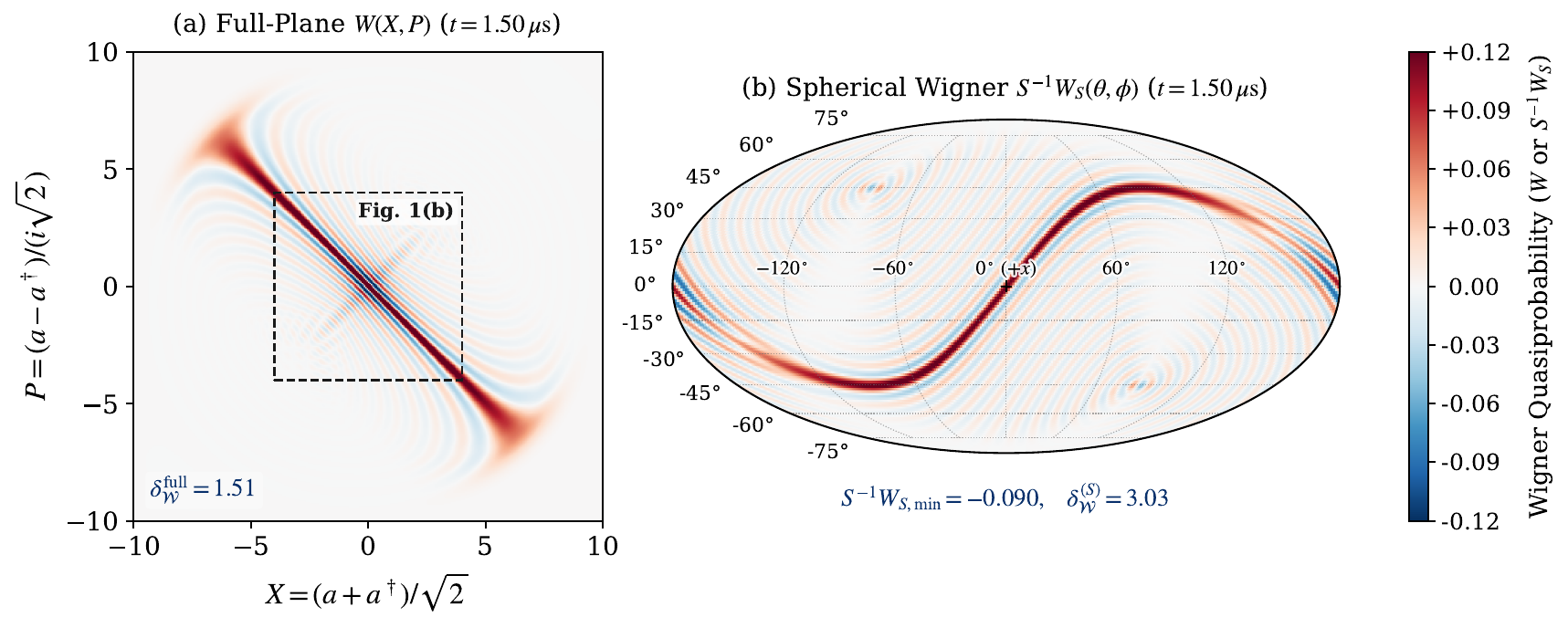}
\caption[]{\revb{Full-support planar and intrinsic $\mathrm{SU}(2)$ spherical Wigner reconstructions of the same open-system $N = 50$ Floquet TAT state ($\chi_{\mathrm{eff}} = 1.0\,\mu\mathrm{s}^{-1}$, $\Gamma_{\mathrm{eff}}/\chi_{\mathrm{eff}} = 5 \times 10^{-4}$, adaptive $\chi_{\max} = 128 \to 256$, $p_{\mathrm{sym}} \approx 0.9903$) at $t = 1.50\,\mu\mathrm{s}$ extracted via the Exact Bilateral DP Dicke Projector. (a) Full-plane Fock-mapped Wigner function $W(X, P)$ over $[-10, 10]^2$, with the dashed square indicating the zoomed $[-4, 4]^2$ window of Fig.~\ref{fig:wigner}(b); capturing the curved high-excitation wings ($\langle\hat{n}\rangle \approx 23.34$) recovers full normalization ($\iint_{[-12,12]^2} W\,dX\,dP = 1.0000$) and increases the integrated negativity volume from $\delta_{\mathcal{W}}^{[-4,4]^2} \approx 0.96$ to $\delta_{\mathcal{W}}^{\mathrm{full}} = 1.508 \approx 1.51$. (b) Mollweide equal-area projection of the intrinsic $\mathrm{SU}(2)$ spherical Wigner function $S^{-1}W_S(\theta, \phi)$ [Eq.~\eqref{eq:Wigner_SU2}] centered at the $+x$ pole ($\theta = \pi/2, \phi = 0$, with longitude $\phi \propto +J_y$ and latitude $\propto +J_z$, accounting for the $90^\circ$ orientation relative to panel (a)), resolving global Bloch-sphere wrapping and antipodal interference fringes ($S^{-1}W_{S,\min} = -0.090$, $S^{-1}W_{S,\max} = 0.139$, spherical negativity volume $\delta_{\mathcal{W}}^{(S)} = 3.03$, exact normalization $\int W_S\,d\Omega = 1.0000$, and conditional purity $\frac{4\pi}{2S+1}\int W_S^2\,d\Omega = \mathrm{Tr}(\rho_S^2) = 0.9962$). Shared color scale saturated at $\pm 0.12$.}}
\label{fig:spherical_wigner}
\end{figure*}

Figure~\ref{fig:wigner} illustrates how $t = t_{\mathrm{opt}}$ serves as the dynamical boundary separating the Gaussian metrological regime from the non\hyph Gaussian \revb{resource} regime:
\begin{itemize}
    \item \textbf{Optimal Gaussian Squeezing (\rev{$t_{\mathrm{opt}}^{\mathrm{open}} = 0.8937\,\mu\mathrm{s}$}):} As shown in Fig.~\ref{fig:wigner}(a) and the solid red curve in Fig.~\ref{fig:wigner}(c), whereas the ideal closed\hyph system Floquet TAT benchmark (dashed blue curve) reaches its theoretical minimum \rev{$\xi_R^2 \approx 0.0722$ ($-11.41\,\mathrm{dB}$) at $t_{\mathrm{opt}}^{\mathrm{ideal}} = 0.8945\,\mu\mathrm{s}$ ($\chi_{\mathrm{eff}}t_{\mathrm{opt}}^{\mathrm{ideal}} = 0.8945$)}~\cite{Kitagawa1993,Liu2011}, local dissipation ($\Gamma_{\mathrm{eff}} = 5 \times 10^{-4}\chi_{\mathrm{eff}}$) \rev{shifts the continuous\hyph time open\hyph system turnaround slightly earlier to $t_{\mathrm{opt}}^{\mathrm{open}} = 0.8937\,\mu\mathrm{s}$ (an advance of $\approx 0.8\,\mathrm{ns}$, verified independently via an exact PIQS solution) while gently raising the squeezing valley to $\xi_R^2 \approx 0.0734$ ($-11.34\,\mathrm{dB}$ below the Standard Quantum Limit, with $\mathrm{Tr}\,\hat{\rho}^2 \approx 0.9877$ and $p_{\mathrm{sym}} \approx 0.9942$) at the nearest $\Delta t = 0.02\,\mu\mathrm{s}$ evolution grid point $t = 0.90\,\mu\mathrm{s}$ ($\chi_{\mathrm{eff}}t = 0.90$), confirming strong robustness of the Floquet TAT squeezing against realistic single\hyph atom relaxation}. For $0 < t \le t_{\mathrm{opt}}$, $\langle\hat{n}\rangle \ll 2S$ \revb{($\langle\hat{n}\rangle \approx 6.14$ at $t = 0.90\,\mu\mathrm{s}$)}, so the quadratic squeezing term $H_{\mathrm{TAT}}^{(0)}$ dominates, yielding a smooth Gaussian\hyph like diagonal ellipse \rev{with only incipient curvature\hyph induced negativity ($W_{\min} \approx -0.029$, \revb{windowed} $\delta_{\mathcal{W}}^{[-4,4]^2} \approx 0.14$\revb{, vs.\ $0.156$ over the full plane in the unitary limit})}.

    \item \textbf{Over\hyph Squeezed Non\hyph Gaussian Interference ($t > t_{\mathrm{opt}}$, benchmarked at \rev{$t = 1.50\,\mu\mathrm{s}$}):} For $t > t_{\mathrm{opt}}$ (shaded region in Fig.~\ref{fig:wigner}(c)), $\xi_R^2(t)$ rebounds upward and crosses \rev{the Standard Quantum Limit ($\xi_R^2 > 1$) at $t \approx 1.372\,\mu\mathrm{s}$ before climbing to $\xi_R^2 \approx 18.20$ at $t = 1.50\,\mu\mathrm{s}$} as the wavepacket wraps around the Bloch sphere \revb{($\langle\hat{n}\rangle \approx 23.34 \approx S$)}. However, as shown in Fig.~\ref{fig:wigner}(b), this \rev{over\hyph squeezed} regime is precisely where finite-$N$ Bloch\hyph sphere curvature dominates. At \rev{$t = 1.50\,\mu\mathrm{s}$ ($\chi_{\mathrm{eff}}t = 1.50$)}, the central positive ridge (\rev{$W_{\max} \approx 0.317$}) is flanked by sharp, alternating positive (red) and negative (deep blue) interference ripples across phase space, with the Wigner minimum plunging to \rev{$W_{\min} \approx -0.154$} and \rev{the} \revb{windowed} negativity volume \revb{over $[-4,4]^2$} reaching \rev{$\delta_{\mathcal{W}}^{[-4,4]^2} \approx 0.96$ (nearly sevenfold above its value at $t_{\mathrm{opt}}$)}. \revb{Because high\hyph excitation tails ($\langle\hat{n}\rangle \approx 23.34$) extend well beyond $[-4,4]^2$ (which captures $\approx 35\%$ of the total quasiprobability normalization at $t = 1.50\,\mu\mathrm{s}$), $\delta_{\mathcal{W}}^{[-4,4]^2} \approx 0.96$ constitutes a conservative lower bound on the full\hyph plane negativity volume Eq.~\eqref{eq:Wigner_volume}, which converges over $[-12,12]^2$ (where $\iint W\,dX\,dP = 1.0000$) to $\delta_{\mathcal{W}}^{\mathrm{full}} = 1.508 \approx 1.51$ for the exact same open\hyph system state [Fig.~\ref{fig:spherical_wigner}(a)]. Furthermore, directly evaluating the intrinsic $\mathrm{SU}(2)$ spherical Wigner function Eq.~\eqref{eq:Wigner_SU2} from the extracted open\hyph system matrix $\rho_{n_L,n_R}$ [Fig.~\ref{fig:spherical_wigner}(b)] resolves the full Bloch\hyph sphere wrapping and antipodal interference around $-x$ without tangent\hyph plane distortion, yielding $S^{-1}W_{S,\min} = -0.090$, $S^{-1}W_{S,\max} = 0.139$, and spherical negativity volume $\delta_{\mathcal{W}}^{(S)} = 3.03$ (with spherical normalization $1.0000$ and purity $\frac{4\pi}{2S+1}\int W_S^2\,d\Omega = \mathrm{Tr}(\rho_S^2) = 0.9962$).}
\end{itemize}

\rev{Notably, the close overall agreement between the open\hyph system trajectory (solid red) and the ideal closed\hyph system benchmark (dashed blue) in Fig.~\ref{fig:wigner}(c) reflects the physical strong\hyph coherence regime of realistic cavity\hyph QED parameters ($\Gamma_{\mathrm{eff}}/\chi_{\mathrm{eff}} = 5 \times 10^{-4}$). Specifically, the single\hyph atom spontaneous\hyph emission probability remains as small as $\Gamma_{\mathrm{eff}} t_{\mathrm{opt}}^{\mathrm{open}} \approx 4.5 \times 10^{-4}$ at optimal squeezing and $7.5 \times 10^{-4}$ at $t = 1.50\,\mu\mathrm{s}$, keeping the many\hyph body purity high ($\mathrm{Tr}\,\hat{\rho}^2 = 0.9771$) and the symmetric\hyph subspace leakage minimal ($1 - p_{\mathrm{sym}} \approx 0.97\%$). On the logarithmic scale, the subtle separation between the two curves is most discernible in the turnaround window around and just beyond the squeezing valley ($t \approx 0.89\text{--}1.15\,\mu\mathrm{s}$): because $\xi_R^2$ reaches its global minimum ($\approx 0.0734$) here, even weak local relaxation carries its highest relative weight against the compressed spin variance and advances the continuous\hyph time turnaround slightly earlier ($0.8937\,\mu\mathrm{s}$ vs.\ $0.8945\,\mu\mathrm{s}$), while the adaptive MPS trajectory stays within $0.07\,\mathrm{dB}$ ($1.7\%$) of the ideal unitary minimum; at later times ($t > 1.25\,\mu\mathrm{s}$), unitary finite\hyph $N$ curvature folding drives $\xi_R^2$ rapidly upward by over two orders of magnitude, overwhelming the weak dissipative noise so that the two curves reconverge. Methodologically, approaching and entering the over\hyph squeezed non\hyph Gaussian regime triggers a rapid growth of Liouville\hyph space operator entanglement as phase\hyph space interference fringes form; we therefore employ an adaptive bond\hyph dimension schedule---maintaining $\chi_{\max} = 128$ across the initial Gaussian squeezing stage ($t \le 0.86\,\mu\mathrm{s}$, where $|\mathrm{Tr}\,\hat{\rho} - 1| < 1.5 \times 10^{-4}$) and doubling the bond dimension to $\chi_{\max} = 256$ ahead of the squeezing valley throughout $0.86\,\mu\mathrm{s} < t \le 1.50\,\mu\mathrm{s}$---which preserves full\hyph space trace conservation ($|\mathrm{Tr}\,\hat{\rho} - 1| < 1.3 \times 10^{-3}$) and monotonic purity decay without spurious truncation\hyph induced damping. As an independent accuracy cross\hyph check at $N = 50$, an exact permutational\hyph invariant master\hyph equation (PIQS) solution~\cite{Shammah2018} confirms that the continuous\hyph time open\hyph system minimum indeed rebounds earlier at $t_{\mathrm{opt}}^{\mathrm{open}} = 0.8937\,\mu\mathrm{s}$ relative to the unitary limit $t_{\mathrm{opt}}^{\mathrm{ideal}} = 0.8945\,\mu\mathrm{s}$ (an advance of $\approx 0.8\,\mathrm{ns}$, such that on the $\Delta t = 0.02\,\mu\mathrm{s}$ grid both discrete minima land on $t = 0.90\,\mu\mathrm{s}$), with an exact dissipative valley of $\xi_R^2 = 0.0726$ ($-11.39\,\mathrm{dB}$, an intrinsic dissipative penalty of $0.02\,\mathrm{dB}$ or $0.5\%$); this confirms that the residual finite\hyph $\chi$ truncation error of our adaptive MPS is only $\approx 1.2\%$ at the squeezing valley and diminishes below $0.5\%$ in $\xi_R^2$ and $0.3\%$ in the deep over\hyph squeezed \revb{windowed} Wigner metrics at $t = 1.50\,\mu\mathrm{s}$ ($W_{\min} = -0.1543$ vs.\ $-0.1544$ and $\delta_{\mathcal{W}}^{[-4,4]^2} = 0.956$ vs.\ $0.954$). Crucially, while $N\Gamma_{\mathrm{eff}}t \ll 1$ keeps the two curves closely aligned at $N = 50$, symmetry\hyph breaking leakage scales as \revb{$p_{\mathrm{sym}}(t) \approx \exp[-\frac{1}{4}(N-1)\Gamma_{\mathrm{eff}}t]$ (i.e., $1 - p_{\mathrm{sym}}(t) \sim \mathcal{O}(N\Gamma_{\mathrm{eff}}t)$)} in larger ensembles, underscoring why a tensor\hyph network framework capable of evolving in the full $4^N$\hyph dimensional Liouville space while tracking $p_{\mathrm{sym}}(t)$ in real time is essential\revb{, particularly when spatial inhomogeneity or disorder breaks permutation symmetry and renders PIQS inapplicable}.}

\begin{figure*}[!t]
\centering
\includegraphics[width=\textwidth]{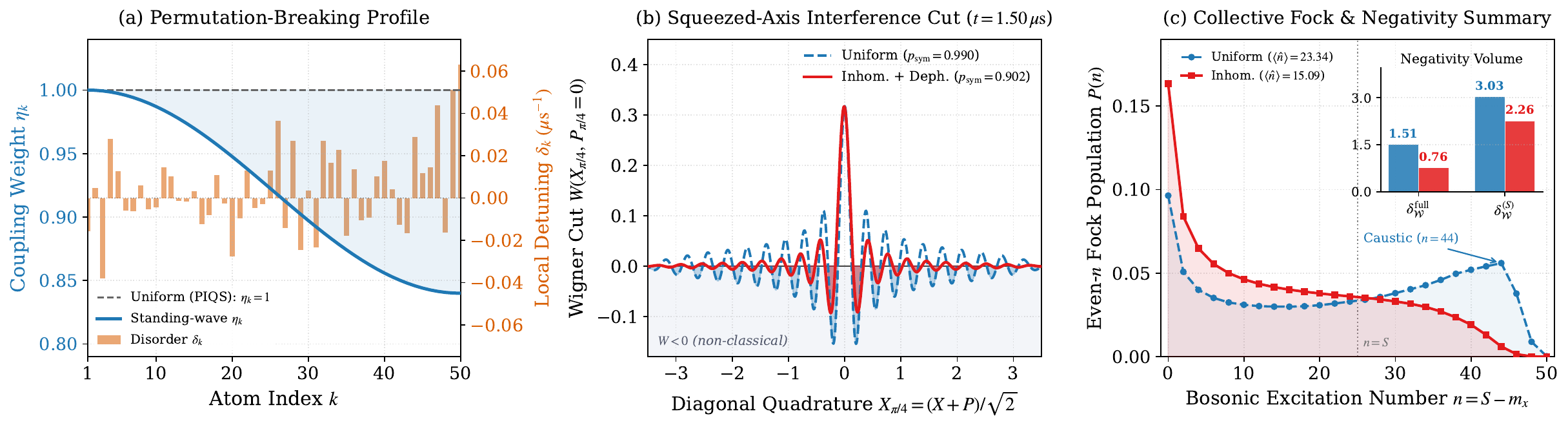}
\caption[]{\revg{Quantitative comparison between the uniform and permutation-breaking inhomogeneous $N = 50$ open-system cavity-BEC benchmarks (beyond PIQS) at $t = 1.50\,\mu\mathrm{s}$, both evolved in the exact $D_{\mathcal{L}} = 6$ FSA Liouvillian MPO ($\chi_{\max} = 128 \to 256$). (a) Site-resolved permutation-breaking profile combining standing-wave coupling modulation $\eta_k = 0.92 + 0.08\cos[\pi(k-1)/(N-1)] \in [0.84, 1.00]$ relative to the uniform PIQS limit $\eta_k = 1$ (left axis; mean pairwise weight $\overline{\eta_i\eta_j} \approx 0.85$) and random local detuning disorder $\delta_k \sim \mathcal{N}(0, 0.02^2)\,\mu\mathrm{s}^{-1}$ (right axis), accompanied by local relaxation and dephasing ($\Gamma_{\mathrm{eff}} = \Gamma_\phi = 5 \times 10^{-4}\chi_{\mathrm{eff}}$). (b) One-dimensional Wigner cross-section $W(X_{\pi/4}, P_{\pi/4} = 0)$ along the squeezed diagonal $X = P$, contrasting the uniform benchmark (dashed blue, $p_{\mathrm{sym}} = 0.990$, $W_{\min} = -0.154$) against the inhomogeneous, dephased benchmark (solid red, $p_{\mathrm{sym}} = 0.902$, $W_{\min} = -0.093$, $|\mathrm{Tr}\,\hat{\rho} - 1| < 2.0 \times 10^{-4}$), with shaded negative interference lobes below the $W = 0$ baseline highlighting persistent non-classicality. (c) Collective even-$n$ Fock population distribution $P(n) = \rho_{n,n}$ (capturing $>99.4\%$ of the conditional weight due to $\Delta n = \pm 2$ parity conservation, with total odd-$n$ leakage $<0.6\%$): the uniform state ($\langle\hat{n}\rangle = 23.34$) develops a finite-$N$ south-pole boundary pile-up caustic at $n = 44$ before plunging to $P(50) \approx 3 \times 10^{-4}$ as $\sqrt{1 - \hat{n}/(2S)} \to 0$, whereas the inhomogeneous state ($\langle\hat{n}\rangle = 15.09$) decays monotonically because its reduced mean coupling slows the over-squeezing advance, keeping the wavepacket short of the south-pole boundary caustic. Inset: closing comparison of the converged full-plane ($\delta_{\mathcal{W}}^{\mathrm{full}}$) and intrinsic spherical ($\delta_{\mathcal{W}}^{(S)}$) Wigner negativity volumes.}}
\label{fig:inhomogeneous_wigner}
\end{figure*}

\paragraph*{\revg{Permutation\hyph Breaking Inhomogeneous Benchmark (Beyond PIQS).}}
\revg{To explicitly demonstrate the computational regime where permutational\hyph invariant solvers~\cite{Shammah2018} are strictly inapplicable, we benchmark an $N = 50$ spatially inhomogeneous cavity\hyph BEC combining standing\hyph wave coupling modulation $\eta_k = 0.92 + 0.08\cos\!\big[\pi(k-1)/(N-1)\big] \in [0.84, 1.00]$ (giving $H_{\mathrm{TAT}}^{\mathrm{inhom}} = \frac{\chi_{\mathrm{eff}}}{N}\sum_{i<j}\eta_i\eta_j(4S_i^z S_j^z - 4S_i^y S_j^y) + \sum_{k=1}^N \delta_k S_k^z$), random site\hyph dependent detuning disorder $\delta_k \sim \mathcal{N}(0, 0.02^2)\,\mu\mathrm{s}^{-1}$, and combined local relaxation and dephasing $\sum_{k=1}^N (\Gamma_{\mathrm{eff}}\mathcal{D}[S_k^-] + \Gamma_\phi\mathcal{D}[S_k^z])$ with $\Gamma_{\mathrm{eff}} = \Gamma_\phi = 5\times 10^{-4}\chi_{\mathrm{eff}}$ [Fig.~\ref{fig:inhomogeneous_wigner}(a)]. Crucially, because $\eta_k$ enters directly as site\hyph dependent weights on the four off\hyph diagonal interaction channels while $\delta_k$ and $\mathcal{D}[S_k^z]$ embed into the $(0 \to \mathrm{final})$ on\hyph site slot, the exact FSA Liouvillian MPO retains its minimal virtual bond dimension $D_{\mathcal{L}} = 6$ (verified to Frobenius error $1.7\times 10^{-15}$ against full $4^N$ exact diagonalization at $N=6$). Using the identical adaptive $W_{\mathrm{II}}$ schedule ($\Delta t = 0.02\,\mu\mathrm{s}$, $\chi_{\max} = 128 \to 256$ for $t > 0.86\,\mu\mathrm{s}$, singular\hyph value cutoff $10^{-7}$), the simulation preserves full\hyph space trace conservation to $|\mathrm{Tr}\,\hat{\rho} - 1| < 2.0\times 10^{-4}$ throughout ($\mathrm{Tr}\,\hat{\rho} = 0.999958$ at $t = 0.90\,\mu\mathrm{s}$ and $0.999807$ at $t = 1.50\,\mu\mathrm{s}$).}

\begin{table}[!t]
\centering
\caption[]{\revg{Quantitative open-system comparison between the uniform and permutation-breaking inhomogeneous $N = 50$ benchmarks at $t = 1.50\,\mu\mathrm{s}$ ($D_{\mathcal{L}} = 6$, $\chi_{\max} = 128 \to 256$). State fidelities $F_{\mathrm{cond}} = \langle\psi_{\mathrm{ideal}}|\rho_S|\psi_{\mathrm{ideal}}\rangle$ and $F_{\mathrm{uncond}} = p_{\mathrm{sym}}F_{\mathrm{cond}}$ are evaluated against the ideal unitary TAT state $|\psi_{\mathrm{ideal}}(1.50\,\mu\mathrm{s})\rangle$ for the uniform benchmark, and ($^\dagger$) against the coupling-rescaled target $|\psi_{\mathrm{ideal}}(t_{\mathrm{eff}} = 1.270\,\mu\mathrm{s})\rangle$ for the inhomogeneous benchmark (unrescaled overlap at $1.50\,\mu\mathrm{s}$: $F_{\mathrm{cond}} = 0.8851$, $F_{\mathrm{uncond}} = 0.7984$).}}
\label{tab:inhom_benchmark}
\footnotesize
\setlength{\tabcolsep}{2.5pt}
\renewcommand{\arraystretch}{1.22}
\begin{tabular*}{\columnwidth}{@{\extracolsep{\fill}} l c c @{}}
\hline\hline
\textbf{Metric ($t = 1.50\,\mu\mathrm{s}$)} & \textbf{Uniform} & \textbf{Inhom.\ + Deph.} \tabularnewline
\hline
Full\hyph space trace $\mathrm{Tr}\,\hat{\rho}$ & $0.9987$ & $0.9998$ \tabularnewline
Symmetry retention $p_{\mathrm{sym}}$ & $0.9903$ & $0.9020$ \tabularnewline
Subspace leakage $1 - p_{\mathrm{sym}}$ & $0.97\%$ & $9.80\%$ \tabularnewline
Conditional purity $\mathrm{Tr}(\rho_S^2)$ & $0.9962$ & $0.9945$ \tabularnewline
Conditional fidelity $F_{\mathrm{cond}}$ & $0.9981$ & $0.9932^\dagger$ \tabularnewline
Unconditional fidelity $F_{\mathrm{uncond}}$ & $0.9884$ & $0.8959^\dagger$ \tabularnewline
Mean excitation $\langle\hat{n}\rangle$ & $23.34$ & $15.09$ \tabularnewline
Planar $W_{\min}/W_{\max}$ & $-0.154/0.317$ & $-0.093/0.315$ \tabularnewline
Windowed $\delta_{\mathcal{W}}^{[-4,4]^2}$ & $0.956$ & $0.558$ \tabularnewline
Full\hyph plane $\delta_{\mathcal{W}}^{\mathrm{full}}$ & $1.508$ & $0.764$ \tabularnewline
Spherical $S^{-1}W_{S,\min/\max}$ & $-0.090/0.139$ & $-0.056/0.185$ \tabularnewline
Spherical $\delta_{\mathcal{W}}^{(S)}$ & $3.03$ & $2.26$ \tabularnewline
\hline\hline
\end{tabular*}
\end{table}

\revg{Table~\ref{tab:inhom_benchmark} and Fig.~\ref{fig:inhomogeneous_wigner}(b)--(c) contrast the resulting open\hyph system observables against the uniform benchmark. At $t = 0.90\,\mu\mathrm{s}$, spatial inhomogeneity and dephasing increase symmetric\hyph subspace leakage to $1 - p_{\mathrm{sym}} = 3.00\%$ ($p_{\mathrm{sym}} = 0.9700$, vs.\ $0.58\%$ uniform), yielding incipient negativity $W_{\min} = -0.0115$ and $\delta_{\mathcal{W}}^{\mathrm{full}} = 0.0542$ ($\delta_{\mathcal{W}}^{[-4,4]^2} = 0.0513$, full\hyph plane normalization $1.00000$). Deep inside the over\hyph squeezed regime at $t = 1.50\,\mu\mathrm{s}$, although leakage reaches $9.80\%$, the conditional symmetric\hyph subspace state maintains high purity ($\mathrm{Tr}(\rho_S^2) = 0.9945$) and resolves pronounced non\hyph Gaussian Wigner negativity across both planar ($\delta_{\mathcal{W}}^{\mathrm{full}} = 0.7644$, normalization $1.00000$) and intrinsic $\mathrm{SU}(2)$ spherical ($\delta_{\mathcal{W}}^{(S)} = 2.26$) reconstructions. Because the ideal unitary TAT target $|\psi_{\mathrm{ideal}}(t)\rangle = e^{-iH_{\mathrm{TAT}}t}|\mathrm{CSS}_x\rangle$ resides entirely within the $J = N/2$ symmetric manifold, our Bilateral DP projection yields exact full\hyph space Uhlmann state fidelities $F_{\mathrm{uncond}} = \langle\psi_{\mathrm{ideal}}|\hat{\rho}|\psi_{\mathrm{ideal}}\rangle = p_{\mathrm{sym}}F_{\mathrm{cond}}$: the uniform benchmark achieves $F_{\mathrm{cond}} = 0.9981$ ($F_{\mathrm{uncond}} = 0.9884$) at $t = 1.50\,\mu\mathrm{s}$, while the inhomogeneous, dephased state retains $F_{\mathrm{cond}} = 0.9932$ ($F_{\mathrm{uncond}} = 0.8959$) against the coupling\hyph rescaled target $|\psi_{\mathrm{ideal}}(t_{\mathrm{eff}} = \overline{\eta_i\eta_j}t = 1.270\,\mu\mathrm{s})\rangle$, showing that simple pulse\hyph duration calibration compensates standing\hyph wave modulation and leaves symmetry leakage as the dominant fidelity budget. Microscopically, the contrast in Fig.~\ref{fig:inhomogeneous_wigner}(c) reflects two competing finite\hyph $N$ effects: in the uniform benchmark, the vanishing Holstein\hyph Primakoff factor $\sqrt{1 - \hat{n}/(2S)} \to 0$ near the antipodal $-x$ pole decelerates the wavepacket into a boundary pile\hyph up caustic at $n = 44$ before plunging to $P(50) \approx 3\times 10^{-4}$ at the hard $\mathrm{SU}(2)$ cutoff; by contrast, standing\hyph wave modulation lowers the mean pairwise coupling to $\overline{\eta_i\eta_j} \approx 0.85$, slowing the over\hyph squeezing advance ($\langle\hat{n}\rangle = 15.09$) so that $P(n)$ decays monotonically without yet reaching the south\hyph pole caustic, while still preserving deep, high\hyph contrast interference fringes across phase space. This confirms both the robustness of curvature\hyph induced non\hyph Gaussianity against realistic cavity inhomogeneity and the unique capability of our $D_{\mathcal{L}} = 6$ MPO $+$ Bilateral DP pipeline beyond permutation\hyph invariant regimes.}

The clear resolution of these high\hyph frequency phase\hyph space ripples demonstrates that our Exact Bilateral DP projection faithfully preserves off\hyph diagonal coherences \rev{$\rho_{m,n}$ ($|m - n| \gg 1$)} from the $4^N$\hyph dimensional Liouville space without suffering from moment\hyph truncation filtering. Scaling this protocol to larger $^{87}\mathrm{Rb}$ condensates follows the universal logarithmic squeezing timescale $\chi_{\mathrm{eff}} t_{\mathrm{opt}} \sim \frac{1}{4}\ln N$~\cite{Lei2026,Liu2011}, where exponential quadrature amplification ($\langle\hat{n}\rangle \sim \frac{1}{4}e^{4\chi_{\mathrm{eff}}t}$) compensates the $\mathcal{O}(1/N)$ bare curvature coefficient to generate macroscopic non\hyph Gaussian interference within logarithmic gate times ($\propto \ln N$). \revb{Physically, preserving deterministic high\hyph retention operation ($1 - p_{\mathrm{sym}}(t) \ll 1$) at macroscopic atom numbers $N$ requires the collective\hyph coherence condition $\Gamma_{\mathrm{eff}}/\chi_{\mathrm{eff}} \ll (N \ln N)^{-1}$ (achievable in high\hyph cooperativity cavities where collective $\chi_{\mathrm{eff}}$ scales favorably with $N$); otherwise, single\hyph atom decay exponentially depletes the $J = N/2$ manifold, highlighting the practical importance of monitoring $p_{\mathrm{sym}}(t)$ directly via our Bilateral DP protocol.}

\section{Conclusion and Outlook}

In summary, we have established a theoretical and computational framework that \revb{bridges} the macroscopic single\hyph cavity neutral\hyph atom $^{87}\mathrm{Rb}$ BEC from a metrological spin\hyph squeezing sensor to a hardware\hyph efficient \revb{deterministic source of continuous\hyph variable non\hyph Gaussian states}, yielding four principal results:
\begin{enumerate}
    \item \textbf{Deterministic Floquet Non\hyph Gaussian State Engineering:} Unlike probabilistic single\hyph photon heralding schemes in atomic ensembles~\cite{McConnell2015} that rely on post\hyph selection, our dual\hyph driven Floquet\hyph engineered TAT protocol~\cite{Lei2026} deterministically synthesizes macroscopic Wigner\hyph negative states via unitary cavity\hyph mediated dynamics.

    \item \textbf{CV Non\hyph Gaussian \revb{Nonlinearity} from the Metrological Over\hyph Squeezed Regime:} Combining a normal\hyph ordered asymptotic Holstein\hyph Primakoff expansion with non\hyph perturbative tensor network simulations, we demonstrated that the over\hyph squeezed regime ($t > t_{\mathrm{opt}}$)---previously treated as a detrimental metrological ceiling in quantum metrology due to the degradation of $\xi_R^2$---harnesses finite-$N$ $\mathrm{SU}(2)$ Bloch\hyph sphere curvature (manifesting at leading order $\mathcal{O}(1/S)$ as \rev{quartic two\hyph photon shearing} $a^\dagger a^3 + (a^\dagger)^3 a$ under pure TAT and exact quartic Kerr $(a^\dagger a)^2$ interactions under asymmetric duty cycles) to synthesize deterministic non\hyph Gaussian states.

    \item \textbf{Truncation\hyph Free $\mathcal{O}(1)$ Liouvillian Automaton MPO:} By formulating our finite\hyph state automaton MPO strictly within the compact $\mathrm{SU}(2)$ spin basis in $4^N$\hyph dimensional Liouville space ($D_{\mathcal{L}} = 6$), we eliminate artificial Fock\hyph space cutoff errors ($N_{\max}$) of circuit\hyph QED simulations~\revb{\cite{Joshi2021}} \revb{and provide a full\hyph space open\hyph system representation that natively extends beyond the permutation\hyph symmetry restriction of PIQS~\cite{Shammah2018} to spatially inhomogeneous cavity couplings and site\hyph dependent dissipation}.
  
    \item \textbf{Decoupled Exact Bilateral Dynamic Programming (DP) Dicke Projection:} By decoupling full\hyph space \rev{$W_{\mathrm{II}}$ Lindblad} evolution from collective phase\hyph space observation, our single\hyph pass Bilateral DP Projector with \revb{step\hyph wise rescaled} log\hyph gamma normalization extracts the $(N+1) \times (N+1)$ symmetric collective subspace projection $\tilde{\rho}_{n_L,n_R}$ without polynomial moment\hyph truncation loss while rigorously tracking the symmetric retention probability $p_{\mathrm{sym}}(t)$, resolving high\hyph frequency non\hyph Gaussian Wigner interference fringes (\rev{$W_{\min} \approx -0.154$, \revb{windowed} $\delta_{\mathcal{W}}^{[-4,4]^2} \approx 0.96$\revb{; converged open\hyph system full\hyph plane $\delta_{\mathcal{W}}^{\mathrm{full}} \approx 1.51$ and intrinsic spherical negativity $\delta_{\mathcal{W}}^{(S)} \approx 3.03$} at $t = 1.50\,\mu\mathrm{s}$ in our converged $N = 50$ benchmark}) \revg{as well as a permutation\hyph breaking inhomogeneous $N = 50$ cavity\hyph BEC benchmark with standing\hyph wave modulation, detuning disorder, and local dephasing yielding $p_{\mathrm{sym}} = 0.9020$, $W_{\min} = -0.093$, and $\delta_{\mathcal{W}}^{\mathrm{full}} = 0.764$ at $t = 1.50\,\mu\mathrm{s}$ with $|\mathrm{Tr}\,\hat{\rho}-1| < 2.0\times 10^{-4}$}, and establishing the cooperativity scaling condition $\Gamma_{\mathrm{eff}}/\chi_{\mathrm{eff}} \ll (N\ln N)^{-1}$ for deterministic macroscopic operation.
\end{enumerate}

Future extensions include \revb{simulating detuning\hyph compensated Kerr cat\hyph state bifurcations ($\chi_{\mathrm{Kerr}} \neq 0$), \revg{incorporating explicit time\hyph dependent Floquet micromotion beyond the leading\hyph order Magnus expansion,} and} coupling multiple cavity modes or spatially addressed $^{87}\mathrm{Rb}$ atomic sub\hyph ensembles to realize continuous\hyph variable entanglement routing and multi\hyph mode bosonic logic gates.

\section*{Acknowledgments}
\textbf{Declaration of AI Assistance:} The author conceived the study, developed the tensor\hyph network and dynamic\hyph programming framework, and performed the numerical simulations. During the preparation of this manuscript, generative AI tools (Google Gemini and DeepSeek) were utilized as computational and editorial assistants to aid with symbolic derivations of the analytical expansions, manuscript drafting and editing, and visualization script generation. The author independently reviewed and verified all mathematical derivations, simulation code, and physical conclusions, and assumes full responsibility for the scientific accuracy, integrity, and originality of this publication.

\section*{\revb{Code and Data Availability}}
\revb{The TeNPy $W_{\mathrm{II}}$ Liouvillian MPO simulation scripts, Exact Bilateral DP Dicke Projector implementation, exact\hyph diagonalization verification scripts, and $N=50$ open\hyph system benchmark datasets supporting the findings of this study are openly available on Zenodo at \url{https://doi.org/10.5281/zenodo.24467817}.}

\appendix

\section{Exact Finite-State Automata for High-Order Spin Multipoles}
\label{app:A}

When evaluating the intrinsic $\mathrm{SU}(2)$ spherical Wigner function $W_S(\theta, \phi)$ via Eq.~\eqref{eq:Wigner_SU2} directly in tensor network form, one requires expectation values of arbitrary collective spin powers $J_\alpha^M = (\sum_{k=1}^N S_k^\alpha)^M$ and mixed correlators $\langle J_x^m J_y^n J_z^k \rangle$.

For a single collective power $J_\alpha^M = (\frac{1}{2}\sum_{k=1}^N \sigma_k^\alpha)^M$, we generalize the finite\hyph state automaton~\cite{Crosswhite2008,McCulloch2007,Hubig2017} to track the cumulative count $s \in \{0, 1, \dots, M\}$ of single\hyph site operator applications across sites $1 \le k \le N$. Because multiple factors in the multinomial expansion of $(\sum_{k=1}^N S_k^\alpha)^M$ can act on the same site $k$, a transition from virtual state $s$ to $s + c$ ($0 \le c \le M - s$) at site $k$ carries the exact binomial weight $\binom{M-s}{c}$:
\begin{align}
\hat{W}_{s,s}^{[k]} &= \mathbb{I}_k \quad (c = 0), \label{eq:W_multipole_0} \\
\hat{W}_{s,s+c}^{[k]} &= \binom{M-s}{c}\left(\frac{\sigma_k^\alpha}{2}\right)^c \nonumber \\
&= \frac{1}{2^c}\binom{M-s}{c}
\begin{cases}
\mathbb{I}_k & (c\text{ even}), \\
\sigma_k^\alpha & (c\text{ odd}),
\end{cases}
\label{eq:W_multipole_c}
\end{align}
where Eqs.~\eqref{eq:W_multipole_0} and \eqref{eq:W_multipole_c} use $(\sigma_k^\alpha)^2 = \mathbb{I}_k$. Contracting $v_L \hat{W}^{[1]} \cdots \hat{W}^{[N]} v_R$ with $v_L = (1, 0, \dots, 0)$ and $v_R = (0, \dots, 0, 1)^T$ (such that the top\hyph right entry $(0, M)$ accumulates the completed product) reproduces $J_\alpha^M$ exactly with bond dimension $D_W = M + 1 = \mathcal{O}(1)$ in $N$:
\begin{align}
\big(v_L \hat{W}^{[1]} &\cdots \hat{W}^{[N]} v_R\big)_{0,M} \nonumber \\
&= \sum_{c_1 + \cdots + c_N = M} \prod_{r=1}^N \binom{M - \sum_{j<r} c_j}{c_r} (S_r^\alpha)^{c_r} \nonumber \\
&= \sum_{c_1 + \cdots + c_N = M} \frac{M!}{c_1! c_2! \cdots c_N!} \prod_{r=1}^N (S_r^\alpha)^{c_r} = J_\alpha^M.
\label{eq:multinomial_proof}
\end{align}
For non\hyph commuting mixed correlators $\langle J_x^m J_y^n J_z^k \rangle$, sequential MPO\hyph MPS application \rev{$(\hat{O}_{\mathrm{MPO}}^{(x,m)}\otimes\mathbb{I})[\allowbreak(\hat{O}_{\mathrm{MPO}}^{(y,n)}\otimes\mathbb{I})[\allowbreak(\hat{O}_{\mathrm{MPO}}^{(z,k)}\otimes\mathbb{I})|\rho\rangle\rangle]]$} with intermediate Singular Value Decomposition (SVD) truncation~\cite{Schollwock2011} avoids forming the product MPO of bond dimension $(m+1)\allowbreak(n+1)\allowbreak(k+1)$ by applying single\hyph axis operators of bond dimension at most $\max(m, n, k) + 1$ per step; however, it still requires $\mathcal{O}(N^2)$ separate evaluations and accumulates intermediate SVD truncation errors, whereas our Bilateral DP Dicke projection in Sec.~\ref{sec:DP} extracts all $(N+1)^2$ symmetric density matrix elements in a single exact pass.

\section{Microscopic Prefactor Calibration\revb{, Step-Wise DP Stability, and Symmetric-Subspace Leakage Rate}}
\label{app:B}

Mapping macroscopic collective spin Hamiltonians onto microscopic 1D tensor networks requires exact tracking of combinatorial prefactors. Expanding $J_z^2 - J_y^2$ into single\hyph site and two\hyph site terms yields:
\begin{align}
J_z^2 - J_y^2 &= \sum_{i=1}^N \big[(S_i^z)^2 - (S_i^y)^2\big] + 2\sum_{i<j} (S_i^z S_j^z - S_i^y S_j^y) \nonumber \\
&= 2\sum_{1 \le i < j \le N} (S_i^z S_j^z - S_i^y S_j^y),
\label{eq:prefactor_calib}
\end{align}
where the single\hyph site terms vanish identically because $(S_i^z)^2 = (S_i^y)^2 = \frac{1}{4}\mathbb{I}$. Consequently, for the collective TAT Hamiltonian $H_{\mathrm{TAT}} = \frac{2\chi_{\mathrm{eff}}}{N}(J_z^2 - J_y^2)$, the microscopic ordered\hyph pair coupling in TeNPy ($H = \sum_{i<j}(c_z S_i^z S_j^z + c_y S_i^y S_j^y)$) must absorb the combinatorial factor of $2$, setting $c_z = -c_y = 4\chi_{\mathrm{eff}}/N$. Omitting this factor of $2$ halves the effective generator ($H \to H/2$), \rev{inducing an exact effective timescale rescaling $t_{\mathrm{eff}} = t/2$ (a twofold temporal dilation)} that our Bilateral DP Dicke projector reproduces to machine precision ($< 10^{-13}$ trace distance against exact diagonalization).

\subsection*{\revb{Exactness of Step-Wise Rescaled Bilateral DP}}
\revb{Let $E_{\alpha_k}^{[k]}(n_L, n_R)$ denote the unrescaled bilateral DP contraction at site $k$. By induction on $k = 0, \dots, N$ using Eqs.~\eqref{eq:DP_recursion}--\eqref{eq:DP_step_rescale} with $\Lambda_0 = 0$ and $\Lambda_k = \sum_{j=1}^k \ln c_j$, the rescaled tensor satisfies $\widetilde{E}_{\alpha_k}^{[k]}(n_L, n_R) = e^{-\Lambda_k} E_{\alpha_k}^{[k]}(n_L, n_R)$ identically while maintaining $|\widetilde{E}_{\alpha_k}^{[k]}(n_L, n_R)| \le 1$ at every step. At $k = N$, combining $\Lambda_N$ with the log\hyph gamma binomial factor inside the exponential of Eq.~\eqref{eq:log_gamma} recovers $\tilde{\rho}_{n_L,n_R}(t) = \binom{N}{n_L}^{-1/2}\binom{N}{n_R}^{-1/2} E_1^{[N]}(n_L, n_R)$ exactly while avoiding both IEEE\hyph 754 overflow of $E_1^{[N]}$ and underflow of $\binom{N}{n}^{-1}$ for arbitrary $N \gg 1029$.}

\subsection*{\revb{Analytic Symmetric-Subspace Leakage Rate}}
\revb{For the initial coherent spin state $|\mathrm{CSS}_x\rangle = |+\rangle_x^{\otimes N}$ under local relaxation $\sum_{k=1}^N \Gamma_{\mathrm{eff}}\mathcal{D}[S_k^-]$, each atom has jump rate $\Gamma_{\mathrm{eff}}\langle +|_x S_k^+ S_k^- |+\rangle_x = \frac{1}{2}\Gamma_{\mathrm{eff}}$ (total rate $\frac{1}{2}N\Gamma_{\mathrm{eff}}$). A single jump $S_k^- |+\rangle_x = \frac{1}{\sqrt{2}}|\downarrow\rangle_z = \frac{1}{2}(|+\rangle_x - |-\rangle_x)$ leaves the $N$\hyph atom state with squared overlap $\frac{1}{2} + \frac{1}{2N}$ onto the $J = N/2$ symmetric Dicke manifold ($|n=0\rangle$ and $|n=1\rangle$), and hence probability $\frac{1}{2}(1 - \frac{1}{N})$ of leaking into the $J = N/2 - 1$ sector. Multiplying by the total jump rate gives the exact initial leakage rate
\begin{equation}
\left.\frac{dp_{\mathrm{sym}}}{dt}\right|_{t=0} = -\frac{1}{4}(N - 1)\Gamma_{\mathrm{eff}},
\label{eq:psym_rate}
\end{equation}
yielding $p_{\mathrm{sym}}(t) \approx \exp[-\frac{1}{4}(N-1)\Gamma_{\mathrm{eff}}t]$ (predicting $p_{\mathrm{sym}} \approx 0.9945$ at $t = 0.90\,\mu\mathrm{s}$ and $0.9908$ at $t = 1.50\,\mu\mathrm{s}$ for $N = 50$, in close agreement with the $W_{\mathrm{II}}$ values $0.9942$ and $0.9903$) and directly establishing the macroscopic collective\hyph coherence condition $\Gamma_{\mathrm{eff}}/\chi_{\mathrm{eff}} \ll (N\ln N)^{-1}$ for logarithmic gate times $\chi_{\mathrm{eff}}t \sim \frac{1}{4}\ln N$.}

\end{document}